\documentclass[a4paper,fleqn,10pt]{article}
\pdfoutput=1
\usepackage{amsmath}
\usepackage{amssymb}
\usepackage{array}
\usepackage{calc}
\usepackage{longtable}
\usepackage{booktabs}
\usepackage{multirow}
\usepackage[T1]{fontenc}
\usepackage{xcolor}
\usepackage{graphicx}
\usepackage{xspace}
\usepackage{units}

\numberwithin{equation}{section}
\usepackage[pdfborder={0 0 0}]{hyperref}
\usepackage[format=hang,labelfont=bf,hypcap=true]{caption}
\usepackage{subcaption}
\usepackage{sectsty}
\usepackage{enumitem}
\usepackage{mciteplus}
\usepackage{preprint/authblk} 
\allsectionsfont{\sffamily}
\subsubsectionfont{\mdseries\itshape\large}
\makeatletter
\DeclareRobustCommand*{\bfseries}{%
  \not@math@alphabet\bfseries\mathbf
  \fontseries\bfdefault\selectfont
  \boldmath
}
\makeatother
\let\spreprint\empty
\newcommand{\preprint}[1]{\def\spreprint{\protect#1}}
\let\sinstitute\empty

\makeatletter
\renewcommand{\maketitle}{\begingroup
  \null\thispagestyle{empty}%
    \ifx\spreprint\empty
      \vskip 5ex
    \else
      \flushright\large\spreprint\vskip 10ex
    \fi
    \vskip 5ex
    \flushleft
      {\sffamily\bfseries\huge\@title}\vskip 6ex
      \@author\vskip 2ex
      \ifx\sinstitute\empty
      \else
        {\small\sinstitute}
      \fi
    \vskip 5ex
  \endgroup
}
\makeatother
\renewenvironment{abstract}{\begin{center}
  {\large\sffamily\bfseries Abstract: }
  \begin{minipage}[t]{0.75\textwidth}
}{\end{minipage}\end{center}\vskip 10ex}

\numberwithin{equation}{section}
\allowdisplaybreaks[2]

\newcommand{\MCatNLO}{M\protect\scalebox{0.8}{C}@N\protect\scalebox{0.8}{LO}\xspace}

\newcommand{\Powheg}{P\protect\scalebox{0.8}{OWHEG}\xspace}

\newcommand{\LOPS}{L\scalebox{0.8}{O}P\scalebox{0.8}{S}\xspace}

\newcommand{\MEPS}{M\scalebox{0.8}{E}P\scalebox{0.8}{S}\xspace}
\newcommand{\NLOPS}{N\scalebox{0.8}{LO}P\scalebox{0.8}{S}\xspace}

\newcommand{\Herwig}{H\protect\scalebox{0.8}{ERWIG}\xspace}

\newcommand{\Pythia}{P\protect\scalebox{0.8}{YTHIA}\xspace}

\newcommand{\Winhac}{W\protect\scalebox{0.8}{INHAC}\xspace}
\newcommand{\Zinhac}{Z\protect\scalebox{0.8}{INHAC}\xspace}

\newcommand{\KKMC}{K\protect\scalebox{0.8}{KMC}\xspace}
\newcommand{\BabaYaga}{B\protect\scalebox{0.8}{ABA}Y\protect\scalebox{0.8}{AGA}\xspace}

\newcommand{\MSbar}{\ensuremath{\overline{\text{MS}}}\xspace}

\newcommand{\Rivet}{R\protect\scalebox{0.8}{IVET}\xspace}

\newcommand{\OpenLoops}{O\protect\scalebox{0.8}{PEN}L\protect\scalebox{0.8}{OOPS}\xspace}

\newcommand{\Sherpa}{S\protect\scalebox{0.8}{HERPA}\xspace}
\newcommand{\Comix}{C\protect\scalebox{0.8}{OMIX}\xspace}

\newcommand{\Amegic}{A\protect\scalebox{0.8}{MEGIC}\xspace}
\newcommand{\CSShower}{C\protect\scalebox{0.8}{SSHOWER}\xspace}

\long\def\symbolfootnote[#1]#2{\begingroup%
\def\thefootnote{\fnsymbol{footnote}}\footnote[#1]{#2}\endgroup}

\newcommand{\im}{\imath}
\newcommand{\jm}{\jmath}
\makeatletter
\providecommand*{\diff}%
{\@ifnextchar^{\DIfF}{\DIfF^{}}}
\def\DIfF^#1{%
\mathop{\mathrm{\mathstrut d}}%
\nolimits^{#1}\gobblespace}
\def\gobblespace{%
\futurelet\diffarg\opspace}
\def\opspace{%
\let\DiffSpace\!%
\ifx\diffarg(%
\let\DiffSpace\relax
\else
\ifx\diffarg[%
\let\DiffSpace\relax
\else
\ifx\diffarg\{%
\let\DiffSpace\relax
\fi\fi\fi\DiffSpace}
\makeatother
\providecommand{\intdphi}[1]{\ensuremath{\int\done\Phi_{#1}\,}}
\providecommand{\zmin}{\ensuremath{z_{\text{min}}}}
\providecommand{\zmax}{\ensuremath{z_{\text{max}}}}

\providecommand{\xmax}{\ensuremath{x_{\text{max}}}}
\newcommand{\ijt}{{\widetilde{\im\hspace*{-1pt}\jm}}}

\newcommand{\ajt}{{\widetilde{a\hspace*{-1pt}\jm}}}

\newcommand{\kt}{{\tilde{k}}}
\newcommand{\at}{{\tilde{a}}}
\newcommand{\bt}{{\tilde{b}}}

\newcommand{\done}{{\rm d}}
\newcommand{\order}{\mathcal{O}}
\newcommand{\alphaS}{\alpha_s}

\newcommand{\alphadip}{\alpha_\text{dip}}

\newcommand{\mc}[1]{\mathcal{#1}}
\newcommand{\mr}[1]{\mathrm{#1}}
\newcommand{\mb}[1]{\mathbb{#1}}
\newcommand{\mf}[1]{\mathbf{#1}}
\newcommand{\dst}{\displaystyle}

\newcommand{\nnb}{\nonumber}

\newcommand{\bea}{\begin{eqnarray}}
\newcommand{\eea}{\end{eqnarray}}
\newcommand{\bi}{\begin{itemize}}
\newcommand{\ei}{\end{itemize}}
\newcommand{\hl}{\vphantom{$\int_A^B$}}

\newcommand{\Hl}{\vphantom{$\int\limits_A^B$}}

\newcommand{\Kop}{\boldsymbol{K}}
\newcommand{\Pop}{\boldsymbol{P}}
\newcommand{\KPop}{\boldsymbol{K\!P}}

\newcommand*{\TeV}{\ensuremath{\text{Te\kern -0.1em V}}}
\newcommand*{\GeV}{\ensuremath{\text{Ge\kern -0.1em V}}}
\newcommand*{\MeV}{\ensuremath{\text{Me\kern -0.1em V}}}
\newcommand*{\keV}{\ensuremath{\text{ke\kern -0.1em V}}}
\newcommand*{\eV}{\ensuremath{\text{e\kern -0.1em V}}}

\newcommand{\Gmu}{\ensuremath{G_\mu}}

\providecommand{\Oangle}[1]{\ensuremath{\langle O\rangle^\mathrm{#1}}}
\newcommand{\wacc}{\ensuremath{w_\text{acc}}}

\newlist{myitemize}{itemize}{3}
\setlist[myitemize]{leftmargin=14em}

\newcolumntype{C}{>{\centering\arraybackslash}p{0.14\textwidth}}

\newlength{\unitcharwidth}
\hypersetup{
  pdfauthor={Lois Flower, Marek Sch\"onherr},
  pdftitle={$e^+e^- \to ZH$ at NLO EW matched to a QED parton shower}
}
\preprint{IPPP/26/24\\MCnet-26-03}
\author[1]{Lois~Flower}
\author[2]{Marek~Sch\"onherr}
\affil[1]{Department of Mathematical Sciences, University of Liverpool, Liverpool L69 3BX, UK}
\affil[2]{Institute for Particle Physics Phenomenology, Department of Physics, Durham University, Durham, DH1 3LE, UK}
\title{$e^+e^- \to ZH$ at NLO EW matched to a QED parton shower}
\begin{document}
\vspace*{10mm}
\maketitle
\vspace*{20mm}
\begin{abstract}
To prepare for the next generation particle collider, likely to be a high-energy
precision-frontier electron-positron machine, theoretical predictions must improve
in tandem. One aspect in which it is necessary to build on the progress made at LEP
and at low-energy $e^+e^-$ colliders is in the modelling of initial-state QED radiation
from leptons. In this paper we combine the \MCatNLO parton shower matching method
with QED resummation methods such as the electron
structure function to obtain an automated, process-independent NLO-matched QED
parton shower. The case of an electron-positron
collider provides a particular challenge to the method due to the integrable
singularity present in the lepton structure function, at variance with QCD PDFs.
We develop new methods to allow a standard dipole parton shower to operate
in the presence of this
singularity. We validate the method by examining its dependence on
infrared parameters and by verifying both the NLO-correctness, and the resummation
properties, of the \MCatNLO prediction. Finally, we present results for the
case of Higgs production in association with an
on-shell $Z$ boson at two proposed FCC-ee energies,
the first such predictions at EW NLO+PS accuracy.
\end{abstract}
\newpage
\tableofcontents
\section{Introduction}
\label{sec:intro}

Modern particle collider experiments, such as those at
the Large Hadron Collider and electron-positron `$b$ factories',
are operating in a precision era in which
the experimental error on many measurements is smaller than the
theoretical error on the Standard Model prediction.
In addition to this success of current experiments, there are a wealth
of proposals for a future electron-positron collider, which would provide
the most precise measurements yet of the Standard Model,
and any physics beyond it.
Among the most promising concepts is the future circular collider,
FCC-ee, which is proposed to run at five different centre-of-mass energies, each
with unique physics and precision goals \cite{FCC:2025lpp}.
This situation presents a challenge to the theory and phenomenology community.

The last 20 years have seen a plethora of new methods for increasing theoretical precision, in many cases in a largely automated and process-independent way.
The \Powheg \cite{Nason:2004rx} and \MCatNLO
\cite{Frixione:2002ik} methods allowed the combination of NLO QCD calculations
with the widely used parton shower method for numerical resummation of soft
and collinear logarithms and description of multiple jet production.
Other NLO+PS matching methods have been developed since, such as
\textsc{KrkNLO} \cite{Jadach:2015mza}, \textsc{UNloPs} \cite{Lonnblad:2012ix},
multiplicative-accumulative matching \cite{Nason:2021xke},
and \textsc{ESME} \cite{vanBeekveld:2025lpz}.
Similarly, the development of the multijet merging method
\cite{Catani:2001cc,Lonnblad:2001iq,Krauss:2002up,Alwall:2007fs,Hoeche:2009rj},
and its NLO extension
\cite{Lonnblad:2012ix,Hoeche:2012yf,Gehrmann:2012yg,Frederix:2012ps,Platzer:2012bs},
allowed the combination of dedicated calculations for different
numbers of QCD jets, vastly improving agreement with data for multijet processes.

In addition to the improved combination of fixed-order results with
resummation, there have recently
been a large number of NNLO QCD predictions for many
relevant processes. In particular, the $ZH$ production process
has attracted recent interest \cite{Ferrera:2017zex,Caletti:2025pot}.
By a simple power counting, it is clear that
for processes involving colour-neutral particles, the NLO EW corrections
are at least as important as the NNLO QCD corrections.
Not only this, but for electron-positron colliders, it is essential to model QED
radiation from the initial states since it changes the centre-of-mass energy
of the collision.
This process is known as radiative return, and must be well-understood to
make precise predictions.
In particular, for the determination of the luminosity of a future
lepton collider using the Bhabha scattering and diphoton measurements,
a relative uncertainty of
$10^{-4}$ (0.1 per mille) will be required \cite{FCC:2025lpp}.

QED radiation at colliders can be categorised into beamsstrahlung,
which is collider-specific and arises from interactions between the beams,
contrasted with beam-independent initial-state radiation which is considered
part of the hard interaction. We only consider the latter here; furthermore,
we are not concerned with measurements which are intended to use charged beams
as a source of photons (photoproduction). Whilst the methods outlined in this
study can be used for photoproduction processes, these come with additional
phenomenological considerations which are not applicable to general
electron-positron scattering.
Hence, the established methods for modelling QED initial-state radiation
are the structure function method \cite{Kuraev:1985hb}
and the Yennie-Frautschi-Suura (YFS) method \cite{Yennie:1961ad}.
The former entails exactly solving the DGLAP equations for an electron or positron
in the leading-logarithmic approximation using the leading-order initial condition.
The latter takes a different approach, exponentiating real and virtual
soft divergences to all orders in QED and leaving the remainder as a
reordered perturbative expansion.
In this way, higher-order terms can be neglected whilst still including 
their logarithmically-enhanced contributions.
Whilst the final-state formulation of YFS for decays has been implemented
in \Winhac/\Zinhac \cite{Placzek:2003zg}, \Herwig
\cite{Hamilton:2006xz} and \Sherpa \cite{Schonherr:2008av}, YFS for initial-state
leptons has only been implemented in
\KKMC \cite{Jadach:1999vf} and in \Sherpa \cite{Krauss:2022ajk}.

In this study, we will instead make use of the structure function, by using it
(analogously to hadronic parton distribution functions) as a basis
for a QED parton shower which we will match
to an NLO electroweak calculation.
Our approach is based on the backward-evolution paradigm \cite{Sjostrand:1985xi}.
We note that an NLO-matched QED parton shower which uses forward evolution
to numerically 
solve the DGLAP equations has existed for some time in the event generator
\BabaYaga \cite{Balossini:2006wc}, which is widely used for predictions at $b$ factories
and at other low-energy $e^+ e^-$ colliders. The NLO-matched parton shower
within \BabaYaga
was recently extended to some $2\to 3$ Born processes, in particular
$e^+ e^- \to \mu^+ \mu^- \gamma$
and $e^+ e^- \to \pi^+ \pi^- \gamma$ \cite{Budassi:2026lmr},
both of which are essential input for the interpretation of data
from low-energy electron-positron colliders and the subsequent
extraction of physical parameters.
Comparisons between the predictions of \BabaYaga and \Sherpa for the 
initial-state matched QED shower, and the YFS prediction,
are left to a future publication.
In ref.~\cite{Flower:2024cpj}, a comparison of the YFS method with the
matched QED shower for final-state leptons was made.
A comparison of \BabaYaga's matched parton shower with other Monte Carlo codes for 
low-energy lepton and pion production was made in ref.~\cite{Aliberti:2024fpq}.
In addition, the general-purpose event generator \Pythia models QED radiation
through a choice of two leading-logarithmic parton showers \cite{Bierlich:2022pfr,
Kleiss:2017iir,Skands:2020lkd};
however, no published results for lepton colliders have been produced
with either the dipole shower or Vincia's antenna approach. The EW angular-ordered shower
in \Herwig is restricted to final-state radiation and non-leptonic initial-state
radiation \cite{Bellm:2019zci,Masouminia:2021kne}.
Both \Pythia and \Herwig, along with many other tools,
implement collinear QED initial-state radiation
using a structure function \cite{Frixione:2022ofv}.

This paper proceeds as follows: in section \ref{sec:methods} we introduce the parton shower, the \MCatNLO method, and the particular considerations relevant for an initial-state shower for $e^+ e^-$; we then present a validation of the methods in section \ref{sec:validation} followed by phenomenological results for $ZH$ production at a future lepton collider in section \ref{sec:results}.
Finally, we offer some concluding remarks in section \ref{sec:conclusions}.

\section{Methods}
\label{sec:methods}

In this section we discuss the construction of a QED dipole shower
that is suitable to describe soft-collinear radiation at $e^+e^-$
colliders, using the backward evolution formulation for initial
state partons that is the \textit{de facto} standard at hadron colliders.
We focus in particular on those initial state radiation aspects that
cannot be directly ported from QED parton showers for hadron
colliders to lepton initial states due to the significantly
different behaviour of the lepton structure function as opposed
to the hadron (e.g.\ proton, neutron, or pion) one.
Final state radiation, on the other hand, follows the familiar
pattern.

\begin{figure}[t!]
\centering
  \begin{subfigure}[b]{0.45\linewidth}
  \includegraphics{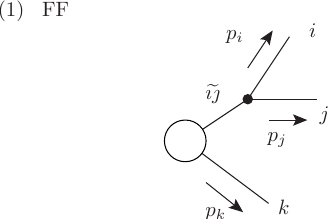}
  \end{subfigure}
  \begin{subfigure}[b]{0.45\linewidth}
  \includegraphics{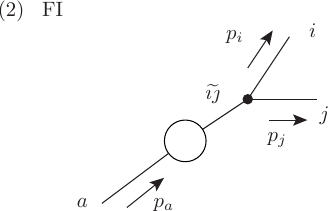}
  \end{subfigure}

  \begin{subfigure}[b]{0.45\linewidth}
  \includegraphics{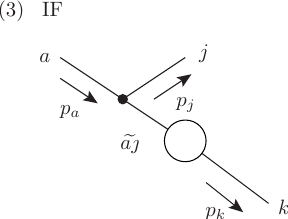}
  \end{subfigure}
  \begin{subfigure}[b]{0.45\linewidth}
  \includegraphics{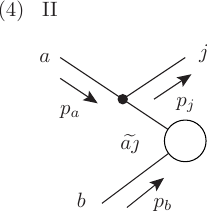}
  \end{subfigure}
  \caption{
    Dipole configurations of the Catani-Seymour splitting functions.
    \label{fig:methods:sf:dips}
  }
\end{figure}

We construct our QED dipole shower in full analogy
to existing QCD parton showers.
In particular, we choose our splitting functions to be the
Catani-Seymour dipoles \cite{Catani:1996vz,Catani:2002hc}
in their QED translation of \cite{Schonherr:2017qcj}.
Thus, following established notation, we will denote a
final-state dipole
splitting process as $\ijt\,(\kt)\to ij\,(k)$,
i.e.\ the splitting of parton\footnote{
  In this paper we denote as \emph{parton} any short-distance
  particle of the model under consideration. In particular,
  in the case of leptons and photons this helps to draw a
  clear distinction between, e.g., long-range beam electrons
  and short-distance partonic electrons that are extracted from
  them through the electron structure function, or, similarly,
  long-range detectable on-shell photons and short-distance
  partonic photons that may still split.
  They are connected through the use of structure (initial state)
  or fragmentation (final state) functions \cite{Kallweit:2017khh}.
} $\ijt$ into partons $i$ and $j$ in the presence of spectator
parton $\kt$, labelled $k$ after absorbing the transverse recoil of
the splitting process.
The corresponding splitting function is
$S_{\ijt(\kt)\to ij(k)}$.
Momenta, masses and other quantities in 
splitting processes are labelled with the splitting index.
Likewise, initial state splittings are denoted as
$a \,(b) \to \ajt^* j \,(\bt)$, with splitting function
$\mr{S}_{a (b) \to \ajt^* j (\bt)}$, where the parton marked
with an asterisk will enter the reduced hard scattering
process. Note that due to the backward-evolution algorithm,
partons `before' and `after' the splitting are reversed in
the case of initial-state splittings.
Depending on whether the emitter, and the spectator, are in
the initial or final state, the dipoles have different forms.
All four different types of dipole are shown in
Fig.\ \ref{fig:methods:sf:dips}.

\subsection{Construction of a QED dipole shower}
\label{sec:method:cssqed}

In implementing a QED parton shower based on Catani-Seymour dipoles,
we follow the construction paradigm of the existing \CSShower,
within \Sherpa, for QCD evolution \cite{Schumann:2007mg}, where first steps towards
the incorporation of some QED evolution effects relevant for hadron
colliders were taken in \cite{Hoeche:2009xc}.
At variance with a QCD dipole shower, however, where the utilisation
of the $N_c\to\infty$ limit results in positive (semi-)definite
splitting functions, no such simplification exists for QED dipoles.
The standard veto algorithm, which forms the basis of all modern
parton and dipole showers \cite{Bengtsson:1986et,Seymour:1994df,
  Sjostrand:2006za}, thus cannot be straightforwardly applied
and its weighted generalisation
\cite{Hoeche:2009xc,Hoeche:2011fd,Platzer:2011dq,Lonnblad:2012hz}
has to be used.

\paragraph*{The standard veto algorithm.} \label{sec:methods:veto}
The veto algorithm is used in parton showers to efficiently
integrate, or sample from, an exponentially decaying
function.
To simplify the notation, we assume the presence of only one
splitting function $f$ without flavour changes, integrated over
all degrees of freedom other than the evolution variable $t$.
The extension to multiple functions, including flavour-changing
splittings, is straightforward.
This splitting function determines the probability $\mc{P}(t,t_0)$ of
a splitting to occur at a scale $t$ if the evolution begins at
a starting scale $t_0$,
\begin{equation} \label{eq:methods:probabilityveto}
  \begin{split}
    \mc{P}(t,t_0)
    \,=&\;
      f(t)\,
      \exp\left[
        -\int_t^{t_0} \done \tilde{t} \, f(\tilde{t})
      \right]
    \,=\;
      \frac{\done\Delta(t,t_0)}{\done t}
    \qquad\text{with}\qquad\!
    \Delta(t_0,t)
    \,=\;
      \exp\left[
        -\int_t^{t_0} \done \tilde{t} \, f(\tilde{t})
      \right]
  \end{split}
\end{equation}
where $\Delta(t,t_0)$ is the Sudakov factor implementing
that no branching occurred during the evolution from $t_0$ to $t$.
Selecting from this distribution is straight forward, provided
the integral of the splitting function, $F$, and its inverse,
$F^{-1}$, are known.
The evolution scales $t$ are then determined according to
\begin{equation}\label{eq:methods:veto:t}
  t = F^{-1}(F(t_0) + \log{\#})\,,
\end{equation}
where $\#$ is a uniformly distributed random number on $[0,1]$.

Unfortunately, in real applications, neither $F$ nor $F^{-1}$
are known.
In its place, an overestimate $g(t) \geq f(t)$ is defined,
where the integral $G$ of the overestimate and its inverse
$G^{-1}$ are known.
We can now propose the new splitting scale $t$ according to
Eq.\ \eqref{eq:methods:veto:t} using $G$ and $G^{-1}$ instead.
This new scale, and the associated splitting, is
accepted with probability $\wacc(t)=f(t)/g(t)$.
If the proposed splitting at scale $t$ is instead rejected,
the evolution continues from this scale, and a new proposal
$\tilde{t}$ is generated.

That this algorithm gives the correct probability of a
single splitting,
$\mc{P}(t,t_0)$, can be explicitly verified.
We compute the probability of a splitting being accepted with
$N$ intermediate rejections,
\begin{equation}\label{eq:methods:veto2}
  \begin{split}
    \mc{P}_N(t,t_0)
    \,=\;&
      \frac{f(t)}{g(t)} \, g(t)
      \exp\left[
        -\int_t^{t_N}\!\! \done \tilde{t}\, g(\tilde{t})
      \right]
      \prod_{i=1}^N
      \left\{
        \int_{t_{i+1}}^{t_{i-1}} \done t_i
        \left( 1-\frac{f(t_i)}{g(t_i)} \right)
        g(t_i)\,
        \exp\left[
          -\int_{t_i}^{t_{i-1}}\!\! \done \tilde{t} \, g(\tilde{t})
        \right]
      \right\} ,
  \end{split}
\end{equation}
and then sum over all possible intermediate rejections.
Identifying $t=t_{N+1}$, and using symmetry under
$t_i \leftrightarrow t_j$ in the product of integrals in the
second term, this gives
\begin{equation}
  \begin{split}
    \sum_{N=0}^\infty \mathcal{P}_N(t,t_0)
    \,=\;
      f(t)\,
      \exp\left[
        -\int_t^{t_0}\!\! \done \tilde{t}\, g(\tilde{t})
      \right]
      \sum_{N=0}^\infty
      \frac{1}{N!}
      \left(
        \int_t^{t_0}\done\tilde{t}\,
        \left[g(\tilde{t})-f(\tilde{t})\right]
      \right)^N
    \,=\;
      \mathcal{P}(t,t_0)\;.
  \end{split}
\end{equation}
This is exactly the probability distribution we sought to
generate.
The evolution of this newly generated state
continues from the scale $t$ as its starting scale.
Once the infrared cutoff, $t_c$ is reached, the evolution 
stops. This scale regularises the infrared divergences of the
splitting functions by delineating the regime of resolved
emissions, with $t>t_c$, from the unresolved emissions
with $t<t_c$.
Through the unitarity of the above veto algorithm,
expressed as probability conservation, the Sudakov
factor $\Delta(t_c,t_0)$
contains the soft-collinear approximation to the
sum of virtual and unresolved emission corrections.

However, in showing that the veto algorithm reproduces the desired
splitting probability, we have made two crucial assumptions.
First, in order to generate the proposed splitting scale $t$,
we need to require $g(t)\ge 0$ on $[t,t_0]$, such that the
overestimated splitting probability is a probability density.
Second, in order for the acceptance/rejection step to work,
we not only need $g(t)\ge f(t)$, but also $1\ge\wacc(t)=f(t)/g(t)\ge 0$.
Hence, this algorithm in its minimal form is unsuitable to
generate negative splitting functions in particular.

\paragraph{The weighted veto algorithm.}
We now seek to lift these restrictions on the veto algorithm
and extend it to splitting functions that are negative
valued, $f(t)<0$
\cite{Hoeche:2011fd,Platzer:2011dq,Lonnblad:2012hz}.
In the context of QED radiation in a dipole picture, this
situation can occur when there are more than two charged
particles in the process. This can be seen explicitly from
the definition of the charge correlator, Eq.\ \eqref{eq:app:charge_correlator}.

The weighted veto algorithm now introduces a second
overestimate $h(t)$, with known integral $H$ and inverse
$H^{-1}$, that is used to generate a proposed
splitting scale $t$ through Eq.\ \eqref{eq:methods:veto:t}
using $H$ and $H^{-1}$.
However, this scale is accepted with $\wacc(t)=f(t)/g(t)$ as before.
With this modification, the probability of accepting a
splitting scale $t$ after $N$ rejections becomes
\begin{equation}
  \begin{split}
    \mc{P}_N(t,t_0)
    \,=\;&
      \frac{f(t)}{g(t)}\,h(t)
      \exp\left[
        -\int_t^{t_N}\!\! \done \tilde{t} \, h(\tilde{t})
      \right]
      \prod_{i=1}^N
      \left\{
        \int_{t_{i+1}}^{t_{N-1}} \done t_i
        \left(1-\frac{f(t_i)}{g(t_i)}\right)
        h(t_i)
        \exp\left[
          -\int_{t_i}^{t_{i-1}}\!\! \done \tilde{t}\,h(\tilde{t})
        \right]
      \right\}\,,
  \end{split}
\end{equation}
again with $t=t_{N+1}$.
We see that we recover the correct expression for
the splitting probability $\mc{P}(t,t_0)$ when 
for every accepted splitting scale $t$, we apply the weight
\begin{equation}\label{eq:methods:veto:weight}
  \begin{split}
    w_N(t,t_N,...,t_1,t_0)
    \,=\;&
      \frac{g(t)}{h(t)}\;
      \prod_{i=1}^N
      \frac{g(t_i)}{h(t_i)}
      \frac{h(t_i) - f(t_i)}{g(t_i) - f(t_i)}\;,
  \end{split}
\end{equation}
where the product is over every rejected scale $t_i$, up to the total
number $N$ of rejected splittings.
This gives
\begin{equation}
  \begin{split}
    w_N \mathcal{P}_N(t,t_0)
    \,=\;&
    f(t)
    \exp\left[
      -\int_{t}^{t_N}\!\! \done \tilde{t} \, h(\tilde{t})
    \right]
    \prod_{i=1}^N
      \left\{
        \int_{t_{i+1}}^{t_{i-1}} \done t_i
        \left( 1-\frac{f(t_i)}{h(t_i)} \right)
        h(t_i)\,
        \exp\left[
          -\int_{t_i}^{t_{i-1}}\!\! \done \tilde{t} \, h(\tilde{t})
        \right]
      \right\}\,,
  \end{split}
\end{equation}
and, after summing over $N$ from zero to infinity, the correct
expression for $\mc{P}(t,t_0)$.

Again, we have made implicit assumptions.
We require $h(t)\ge 0$ on $[t,t_0]$, such that the overestimated
splitting function is still a probability density.
In addition, $0 \le f(t)/g(t) \le 1$ must hold so that the
accept/reject step is probabilistically meaningful.
In practice, in cases where the charge correlator is the
only source for a negatively-valued splitting function, we
choose $h(t)$ as the traditional overestimate for $|f(t)|$
and $g(t)=-h(t)$.
With this choice the acceptance weight is always $-1$ and
each rejection weight is strictly positive.
However, with the weighted veto algorithm we no longer have the 
requirement that $g(t)$ must be analytically integrable ---
it is not used until the accept/reject step ---
so it can be more efficient to choose $g(t)=cf(t)$. With this option,
to avoid large or diverging rejection weights where $f(t)$ and $g(t)$
coincide, we choose $c \approx 2$. We use this option in
the \MCatNLO matched emission.

\paragraph*{Evolution variable.}
The choice of the evolution variable $t$ is one of the defining
properties of any parton shower algorithm as it defines the
quantity that is resummed in its Sudakov factor.
Here we follow similar choice made in the QCD showers,
which generally fall under the category of transverse 
momentum ordered showers \cite{Schumann:2007mg}, but to which 
various modifications have been made \cite{Carli:2010cg,Hoeche:2014lxa,ATLAS:2021yza}.
For initial-state splittings $a \,(b) \to \ajt^* j \,(\bt)$,
we distinguish two cases, choosing the modified transverse momentum
$\bar{k}_\mr{T}^2$ if a soft-photon
singularity is present in the splitting function,
and the reduced virtuality $\bar{q}^2$ otherwise,
defined as
\begin{equation} \label{eq:methods:IIkTordering}
  \begin{split}
    t_{f\to f^*\gamma}
    \,=\;&
      \bar{k}_\mr{T}^2 = (Q^2-m_f^2-m_b^2)\,\frac{y}{z}\,(1-z) \\
    t_{f\to \gamma^*f}
    \,=\;&
      \bar{q}^2 = (Q^2-m_f^2-m_b^2)\,\frac{y}{z}\\
    t_{\gamma\to f^*\!\bar{f}}
    \,=\;&
      \bar{q}^2 = (Q^2-2m_f^2-m_b^2)\,\frac{y}{z} + 2m_f^2 \,.
  \end{split}
\end{equation}
Likewise, for final state splittings $\ijt\,(\kt)\to ij(k)$ we
distinguish two cases,
\begin{equation} \label{eq:methods:FFkTordering}
  \begin{split}
    t_{f\to f\gamma}
    \,=\;&
      \bar{k}_\mr{T}^2 = (Q^2-m_f^2-m_k^2)\,y\,(1-z)
      \\
    t_{\gamma\to f\!\bar{f}}
    \,=\;&
      \bar{q}^2 = (Q^2-2m_f^2-m_k^2)\,y + 2m_f^2\,,
  \end{split}
\end{equation}
In the above, $Q^2$ is the dipole invariant mass,
$z$ is the light-cone momentum fraction retained
by the emitting parton in the splitting process,
and $y$ is the remaining dipole variable.
Their precise definitions depend on the
dipole type and are given in App.\ \ref{app:sfs}.

The use of the reduced virtuality in photon splittings further
resums higher-order logarithms originating in the
QED $\beta$-function\footnote{In the corresponding QCD shower,
which may be interleaved with the QED shower, the same choice is made for all
gluon splittings.} \cite{Brodsky:1982gc}.
Similar arguments apply to the choice of the value of the
coupling $\alpha$, which we choose to be its effective coupling evaluated
at the respective photon virtuality.%
\footnote{Please note, since $\alpha$ is commonly
renormalised in the on-shell scheme \cite{Denner:2019vbn} the argument
of the effective coupling $\alpha(t)$ plays a different role than
the argument of the commonly $\MSbar$-renormalised strong coupling
$\alphaS(\mu_{\mr{R}})$ evaluated at the renormalisation scale
$\mu_{\mr{R}}$.
Other parton showers make different choices in this regard \cite{Bahr:2008pv,Bierlich:2022pfr}.}
Consequently, photon emissions couple with $\alpha(0)$,
since the photon wavefunction renormalisation cancels the
light-fermion logarithms which appear. Photon splittings, on the
other hand, couple with $\alpha(t')$ for a splitting at scale $t'$.
Where a photon was produced by a parton shower emission at scale $t$
and later splits at scale $t'$, its coupling should be retrospectively
rescaled by $\alpha(t\ensuremath{'})/\alpha(0)$.

These choices, together with the splitting functions detailed below,
ensure the correct coherence in the soft-photon limit, and
angular ordering of successive emissions where appropriate.
We note at this point that, since we restrict ourselves to the
use of an electron structure function only (see
Sec.\ \ref{sec:methods:electronSF}), the initial state photon density
is strictly zero.
In consequence, only the splittings that use the first
definition of the evolution variable in Eq.\ \eqref{eq:methods:IIkTordering},
$t_{f\to f^*\gamma}$
are present in the results presented in this paper.

\paragraph*{Infrared cutoff.}
In QCD showers, the infrared cutoff $t_c$ of the perturbative
evolution is dictated by the transition from
perturbative QCD, where quarks and gluons are the relevant
degrees of freedom, to non-perturbative QCD and the colour-confined
regime, where hadrons are the physical objects.
Therefore, $t_c$ is usually placed at scales
around 1\,GeV and is subject to the tuning of the non-perturbative
models \cite{Buckley:2011ms,Bierlich:2022pfr,Chahal:2022rid}.
Conversely, QED is infrared free and perturbative throughout---leptons
and photons remain the
relevant physics objects at long distance.
Hence, the IR cutoff needs only to play its role as IR
regulator and must be set low enough to not interfere with
physical processes of interest, including soft-photon emission
and photons splitting into electron-positron pairs.
We, thus, set the default IR cutoff for splittings involving
only leptons and photons to a value near the electron mass.
Nonetheless, the IR cutoff for photon emissions off
quarks and photons splitting into quark-antiquark pairs
remains at the order of 1\,GeV, in order to not interfere
with their hadronisation process.
We will investigate the possible choices of the IR
cutoff in Sec.\ \ref{sec:validation}.

\paragraph*{Splitting functions and kinematics.}

Finally,
we introduce the parton shower splitting functions that recreate
the soft and collinear limits of matrix elements with
one additional emission.
We start by considering the soft-collinear limit of the
$(n+1)$--particle partonic cross section in terms of the
$n$--particle partonic cross section,
\begin{equation}\label{eq:methods:SF:approx-corr}
  \done\hat{\sigma}_{n+1}
  =
    \done\hat{\sigma}_n \otimes
    \sum_{\{\ajt,\bt\}}
    \sum_\mathrm{f}
    \done t\,
    \done z\,
    \frac{\done \phi}{2\pi}\;
    \mf{S}_{a(b)\to \ajt \,\mr{f}\, (\bt)}(t,z,\phi)\;.
\end{equation}
Therein, $\mf{S}_{a(b)\to \ajt \,\mr{f}\, (\bt)}(t,z,\phi)$ is the
matrix-valued (in spin/polarisation space) splitting operator
\cite{Dittmaier:1999mb,Dittmaier:2008md,Schonherr:2017qcj} retaining the
full dependence on the azimuthal splitting angle $\phi$,
and $\mathrm{f}$ runs over possible flavours of the emitted
parton (in QED, $\mathrm{f}=q,\ell,\gamma$).
The $\otimes$-operator is defined in full analogy to refs.
\cite{Catani:1996vz,Catani:2002hc,Dittmaier:2008md,Schonherr:2017qcj}.
Integrating out the dependence on the spin/polarisation
of the emission, we arrive at the spin-averaged splitting
function
$\mr{S}_{a(b)\to \ajt \,\mr{f}\, (\bt)}=\langle\mf{S}_{a(b)\to \ajt \,\mr{f}\, (\bt)}\rangle$,
\begin{equation}\label{eq:methods:SF:approx-av}
  \langle\done\hat{\sigma}_{n+1}\rangle
  =
    \done\hat{\sigma}_n \cdot
    \sum_{\{\ajt,\bt\}}
    \sum_\mathrm{f}
    \done t
    \done z\;
    \mr{S}_{a(b)\to \ajt \,\mr{f}\, (\bt)}(t,z)\;.
\end{equation}
As $\mr{S}_{a(b)\to \ajt \,\mr{f}\, (\bt)}$ is now a scalar quantity,
the $n$-parton cross section now factorises completely,
as a simple multiplication of two scalar numbers.
The real-valued splitting function $\mr{S}_{a(b)\to \ajt \,\mr{f}\, (\bt)}$,
reduces further to the familiar unregularised
Altarelli-Parisi splitting kernel $\hat{P}_{a\mr{f}}$
in the collinear limit and, for $\mr{f}=\gamma$
only, the (partial-fractioned) eikonal in the soft limit
\cite{Catani:1996vz,Dittmaier:1999mb,Catani:2002hc,
  Dittmaier:2008md,Schonherr:2017qcj},
\begin{equation}
  \mr{S}_{a(b)\to \ajt \,\mr{f}\, (\bt)}(t,z)
  \xrightarrow{p_a||p_\mr{f}}
  \frac{8\pi\,\alpha}{z\,2p_ap_\mr{f}}\,\hat{P}_{\ajt a}(z)
  \quad\text{and}\quad
  \mr{S}_{a(b)\to \ajt \gamma (\bt)}(t,z)
  \xrightarrow{{p_\gamma\to\lambda q}\atop{\lambda\to 0}}
  \frac{8\pi\,\alpha}{\lambda^2\,2p_aq}\, Q_a Q_b\,
  \frac{p_ap_b}{(p_a+p_b)q}
  \,.\hspace*{-20mm}
\end{equation}
The choice of $\alpha$ in the above expressions is
discussed in Sec.\ \ref{sec:methods:ewscheme}.
The precise forms of all splitting functions are detailed in
App.\ \ref{app:sfs}. However, before discussing the relevant
changes to the method which are necessary for initial-state
leptonic evolution, we write the relevant splitting function
here. For massless initial-state fermions, the splitting
function we use for photon emission is\footnote{This splitting
function is modified from the original Catani-Seymour by the replacement
$z \to z+y$ \cite{ATLAS:2021yza}.}
\begin{equation}\label{eq:methods:sf}
  \begin{split}
    \mr{S}_{f(\bt)\to f^*\gamma(b)}(t,z)
    \,=\;
    \frac{8\pi\,\alpha}{2p_ap_j}\,
    \frac{1}{z}\, Q_f Q_b
    \left[\frac{2\,(z+y)}{1-z} + (1-z-y) \right]
    \,,
  \end{split}
\end{equation}
where $Q^2$ is the dipole invariant mass,
$z$ is the light-cone momentum fraction retained
by the emitting parton in the splitting process,
$y$ is the remaining dipole variable (see App.\ \ref{app:sfs}
for their definitions in terms of momenta).
Finally, taking into account the initial state
partonic electrons emerged from a composite beam
particle, we absorb the change in structure function
or PDF into our splitting function
\begin{equation}\label{eq:methods:sf_fratio}
  \begin{split}
    \overline{\mr{S}}_{a(b)\to \ajt \,\mr{f}\, (\bt)}(t,z)
    \,=\;
    \mr{S}_{a(b)\to \ajt \,\mr{f}\, (\bt)}(t,z)\;
    \frac{f_a(\frac{x}{z},t)}{f_\ajt(x,t)}
    \,,
  \end{split}
\end{equation}
where $x$ is the momentum fraction of the original
parton $\ajt$ and $f_\ajt(x,t)$ is its PDF or structure
function, while $f_a(\tfrac{x}{z},t)$ is the PDF or structure
function of the newly produced parton $a$.
Currently, the description of leptonic structure in Monte Carlo
tools is restricted to a structure function which does not encode the
possibility of extracting a photon from the lepton, rather than a full
parton distribution function description.\footnote{For the
processes we consider here, which are driven by large-$x$ initial states,
the contribution of the photonic initial state can be estimated as
$f_\gamma^e/f_e^e \sim 1-x$ \cite{Bertone:2019hks},
additionally mediated by the splitting probability $\mr{S}_{\gamma \to ee}$,
resulting in a sub-per-mille correction.
}
Therefore, the only dipoles
we need to consider specifically for initial-state radiation
at lepton colliders are these
photon-emission dipoles.

For the kinematics, since we want to match our shower to an exact NLO EW
calculation,
we choose the original dipole kinematics of
\cite{Catani:2002hc,Schumann:2007mg} while noting
that for pure showering the alternative kinematics
of \cite{Hoeche:2009xc} are also implemented.

With these splitting functions defined, we can now use
the above veto algorithm to improve our leading order
accurate infrared-safe observable $O$, described
using the LO squared matrix element $\mr{B}$ on the
leading order phase-space $\Phi_n$, to include the effect
of soft-collinear emissions.
The LO plus parton shower prediction for $O$ is then
\begin{equation}\label{eq:methods:LOPS}
  \begin{split}
    \Oangle{\text{\LOPS}}
    \,=\;
      \intdphi{n}
      \mr{B}(\Phi_n) \, \mathcal{F}_n(t_n,O),
  \end{split}
\end{equation}
where the functional
\begin{equation}\label{eq:methods:PSfunc}
  \begin{split}
    \mc{F}_n(t_n,O)
    \,=\;&
      \Delta_n(t_c,t_n) \, O(\Phi_n)
      +
      \sum_{\{\ajt,\bt\}} \sum_{\mr{f}}
      \int_{t_c}^{t_n}\hspace*{-5pt}\done t
      \int_{\zmin}^{\zmax}\hspace*{-10pt}\done z\;
      \overline{\mr{S}}_{a(b)\to \ajt \,\mr{f}\, (\bt)}(t,z)\,
      \Delta_n(t,t_n) \,
      \mc{F}_{n+1}(t,O)
  \end{split}
\end{equation}
implements the parton shower evolution,
recursively progressing from the starting scale, $t_n$
on the $n$-parton configuration until the (global) IR cutoff
scale $t_c$ is reached.
The boundaries of the $z$ integral, $[\zmin,\zmax]$,
are given by momentum conservation \cite{Schumann:2007mg}.
The parton shower Sudakov form factor $\Delta(t,t')$, given by
\begin{equation}
  \begin{split}
    \Delta_n(t,t')
    \,=\;&
      \exp\left[
        \sum_{\{\ajt,\bt\}} \sum_{\mr{f}}
      \int_{t}^{t'}\hspace*{-5pt}\done\tilde{t}
      \int_{\zmin}^{\zmax}\hspace*{-10pt}\done\tilde{z}\;
        \overline{\mr{S}}_{a(b)\to \ajt \,\mr{f}\, (\bt)}(\tilde{t},\tilde{z})\,
      \right],
  \end{split}
\end{equation}
resums the ensuing large logarithms.

\paragraph*{Efficiency and dipole identification.}
In the case where many charged particles are present in the
parton shower evolution, considering all dipoles as possible
photon emitters prohibitively impacts the numerical efficiency
of the algorithm due to the consequent abundance of negatively
weighted events.
In particular, a photon, or gluon, of a configuration with
$n$ positively-charged and $m$ negatively-charged
external particles, which splits into a fermion-antifermion
pair will create an additional $2n+2m+2$ opposite-sign charged and
$2n+2m$ like-sign charged dipoles.

In practice, due to the separation in scales in our strongly-ordered
parton shower, and the fact that the intermediate
photon or gluon is neutral, $2n+2m$ of these dipoles pairwise
largely cancel, leaving the $2$ newly created dipoles in the
fermion-antifermion system effectively separated from the dipoles of the
system that emitted the photon or gluon.
Likewise, since these $f\bar{f}$ pairs are usually produced
near-collinearly, soft photon emissions will not resolve their
individual charges if emitted outside the cone spanned by
the dipole; this is the QED analogue of angular ordering.

The solution we adopt in our QED parton shower is to identify
as spectator only the opposite-sign charge same-flavour particle
which results in the smallest dipole invariant mass\footnote{
  This scheme is almost identical to that implemented in
  Vincia \cite{Skands:2020lkd}, and is similar to the one in PYTHIA,
  where new $f\bar{f}$ pairs are tracked manually \cite{Bierlich:2022pfr}.
}.
We have verified for a range of processes that all relevant
distributions are statistically in agreement between the unaltered
and the efficiency-improved shower.
In addition, the first-emission generator which forms
part of the \MCatNLO algorithm is independent of the standard shower;
thus, all spectators are considered in the matched emission even when
the single-spectator mode is selected for the parton shower.

\subsection{The electron structure function}
\label{sec:methods:electronSF}

As QED is perturbative in its scale evolution, an analytic solution
to the DGLAP equations can be obtained at leading-logarithmic
accuracy\footnote{An approach to obtain NLL accurate structure functions
was provided in \cite{Frixione:2019lga}, while the respective
NNLO fixed-order expressions were published in
\cite{Stahlhofen:2025hqd,Schnubel:2025ejl}.}.
This is the electron structure function $f_e(x,Q^2)$, where
$x$ is the momentum fraction carried by the
short-distance electron and $Q^2$ is the scale at which
the long-distance beam electron is probed, usually the
hard scale of the process.
Using the LO initial condition $f_e(x,m_e^2)=\delta(1-x)$
\cite{Skrzypek:1990qs},
it takes the leading-logarithmic (LL) form \cite{Kuraev:1985hb}
\begin{equation}\label{eq:methods:electronSF}
    f_e(x,Q^2)
    =
      \beta\,
      \frac{\exp{\left(
                  -\gamma_e \beta + \frac{3}{4}\beta_S
                 \right)}}
           {\Gamma(1+\beta)}\,
      (1-x)^{\beta-1}
      +
      \sum_{n=1}^\infty \beta_H^n \mathcal{H}_n(x).
\end{equation}
Herein, $\beta = \frac{\alpha}{\pi} \left(\ln{(Q^2/m_e^2)}-1\right)$.
The soft photon residue $\beta_S$ and the hard prefactor $\beta_H$
can be chosen either to equal $\beta$,
or $\frac{\alpha}{\pi} \ln{(Q^2/m_e^2)}$, independently.
In this paper we will choose $\beta_S=\beta_H=\beta$.
While \Sherpa includes the hard coefficients $\mathcal{H}_n$
up to $n=3$, only the leading $n=1$ term is included in this study.

\begin{figure}
  \centering
  \includegraphics[width=0.75\textwidth]{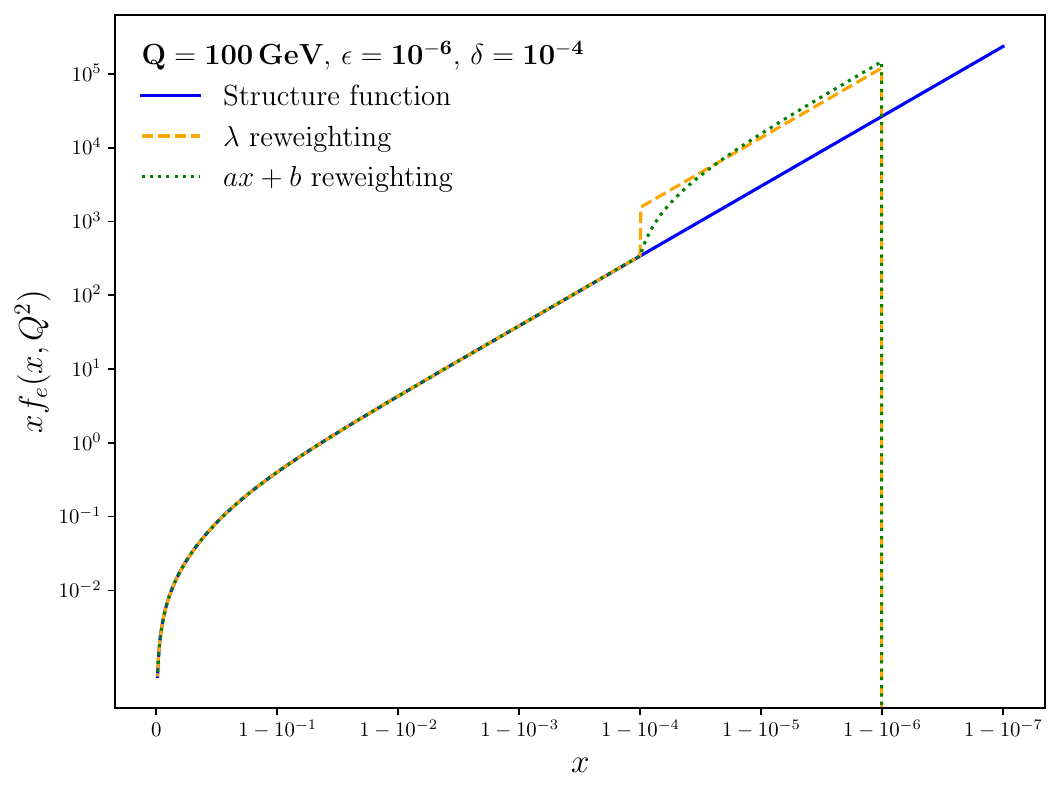}
  \caption{The electron structure function near $x=1$, comparing different
  rescaling schemes for representative choices of $\delta=10^{-4}$,
  $\epsilon=10^{-6}$. Both the $x$ and $y$ axes are logarithmic.
  \label{fig:methods:electronSF}}
\end{figure}

This structure function, like the $\delta$-function it
evolved from, contains an \emph{integrable} singularity at $x=1$.
For event generation, or numerical integration in general,
the presence of this singularity poses an irreconcilable
problem: phase space points within an infinitesimal distance
of $x=1$ will be assigned a infinitely large weight,
overwhelming any numerical precision on the resulting contribution.
Na\"ively introducing a cutoff $\xmax<1$ excludes
an important part of the spectrum and results in a severe
underestimate of the total cross section.
Thus, the standard solution to this problem
is to reweight a portion of the structure function.

Starting from the $x \to 1$ limit of the structure function
of Eq.\ \eqref{eq:methods:electronSF},
\begin{equation}\label{eq:methods:electronSF_limit}
    f_e(x,Q^2)\;
    \stackrel{x\to1}{{}-\!\!\!-\!\!-\!\!\!\longrightarrow}\;
    \beta(1-x)^{\beta-1},
\end{equation}
the minimalist solution \cite{Sjostrand:2006za,Jadach:privatecomms}
is to reweight the original structure function by a constant
in an interval $[1-\delta,1-\epsilon]$, where $\epsilon,\delta\ll1$
and $\epsilon<\delta$,
\begin{equation}
  \begin{split}
    W_e^\lambda(x,Q^2) =
    \begin{cases}
        f_e(x,Q^2) & 0 \leq x \leq 1-\delta \\
        \lambda \,f_e(x,Q^2) & 1-\delta < x \leq 1-\epsilon \\
        0 & \text{else}
    \end{cases}
    \qquad\qquad\text{with}
    \qquad\qquad
    \lambda
    = \frac{\delta^\beta}
           {\delta^\beta - \epsilon^\beta}\;,
  \end{split}
\end{equation}
where the constant $\lambda$ is defined through demanding
\begin{equation}\label{eq:methods:Wlambda_cond}
  \begin{split}
    \int_{1-\delta}^1 \done x\, W_e^\lambda(x,Q^2)
    &=
    \int_{1-\delta}^1 \done x\, f_e(x,Q^2)\;.
  \end{split}
\end{equation}
This solution preserves the total and differential
cross sections as long as $\epsilon\ll\delta$ and $\delta$
is small compared to the typical resolution of the
observable being calculated.\footnote{
  The relation in Eq.\ \eqref{eq:methods:Wlambda_cond} only
  results in the preservation of (differential) cross sections
  if the observable does not have significant $x$-dependence
  over the range $[1-\delta,1]$.
}
However, it is important to note that
the resulting rescaled structure function $W_e^\lambda$
is no longer continuous nor does it satisfy the DGLAP equation.
Its discontinuity at $x=1-\delta$ is of phenomenological
importance once the momenta of the incident short-distance
electrons are changed, e.g.\ through an initial-state parton
shower. On the other hand, its violation of the DGLAP equation
is expected to have negligible impact.

A solution that addresses both issues is to condense the
cross section in $(\epsilon,1]$ in a component proportional to
$\delta(1-x)$,
\begin{equation}\label{eq:method:WC_cond}
  \begin{split}
    W_{e}^{C}(x,Q^2)
    =
      f_e(x,Q^2)\,\Theta(x\leq1-\epsilon)
      \;+\; C\, \delta(1-x)
    \qquad\qquad
    \text{with}
    \qquad\qquad
    C=\epsilon^\beta\;,
  \end{split}
\end{equation}
where $C$ is determined akin to Eq.\ \eqref{eq:methods:Wlambda_cond}.
While continuity and adherence to the DGLAP equation
for $x\in[0,1-\epsilon]$ is manifest in this approach
and, consequently, no artifacts in initial state
parton shower evolution are expected,
for the case of $e^+ e^-$ collisions this
results in complications in event generation as
the phase space is split into four distinct and
discontiguous regions,
\begin{equation}
  (\text{i})\,    x_1\leq1-\epsilon,\, x_2\leq1-\epsilon; \quad
  (\text{ii})\,   x_1\leq1-\epsilon,\, x_2=1; \quad
  (\text{iii})\,  x_1=1,\, x_2\leq1-\epsilon; \quad
  (\text{iv})\,   x_1=1,\, x_2=1\,.
\end{equation}
Thus, we pursue a different ansatz here and define
\begin{equation}
  \begin{split}
    W_{e}^{ax+b}(x,Q^2) =
    \begin{cases}
        f_e(x,Q^2) & 0 \leq x \leq 1-\delta \\
        (ax+b) \,f_e(x,Q^2) & 1-\delta < x \leq 1-\epsilon \\
        0 & \text{else,}
    \end{cases}
  \end{split}
\end{equation}
where
\begin{equation}
  a
  =
    \frac{
      \left(1+\beta\right)\;\epsilon^\beta
    }{
      \delta\,\delta^\beta
      -\left(\delta+\beta(\delta-\epsilon)\right)\epsilon^\beta
    }
  \qquad\qquad\text{and}\qquad\qquad
  b
  = \frac{
      \delta\,\delta^\beta
      -\left(1+\beta(1-\epsilon)\right)\epsilon^\beta
    }{
      \delta\,\delta^\beta
      -\left(\delta+\beta(\delta-\epsilon)\right)\epsilon^\beta
    }
\end{equation}
are determined by demanding equivalence, as in
Eq.\ \eqref{eq:method:WC_cond}, and continuity
at $x=1-\delta$, respectively.
For our later application in our parton shower, this approach
is sufficient to have all ingredients well-defined, when an
emission crosses this threshold.

We illustrate the functional form of all three modifications
of the electron structure function in
Fig.\ \ref{fig:methods:electronSF} and
demonstrate their stability in Tab.\ \ref{table:methods:xs}
and Tab.\ \ref{table:methods:xsVaryDelta}.
Varying the technical parameters $\epsilon$ and $\delta$
across a wide range for different process at various
centre-of-mass energies at electron-positron colliders,
we find stable results if $\epsilon$ and $\delta$ are
chosen not too large, typically $\epsilon\leq 10^{-6}$
and $\delta\leq 10^{-4}$.
If $\delta$ is chosen too small,
i.e.\ too close to $\epsilon$, the results are statistically
compatible but the uncertainty is inflated as the weighting
factor, $\lambda$ or $a$, becomes very large.

\begin{table}[b!]
\centering
\begin{tabular}{lcrcccc}\toprule
\textbf{Process} & \textbf{Rescaling}&& \multicolumn{4}{c}{\textbf{Cross section (pb)}} \\
&&$\epsilon$ & $10^{-9}$&$10^{-8}$&$10^{-7}$&$10^{-6}$\\
\midrule
$e^+e^- \to \nu_\mu \bar{\nu}_\mu$ &$\lambda$&& $2869.0\pm1.6$ & $2869.2\pm1.3$ & $2870.5\pm1.1$ & $2869.7\pm1.0$ \\
91.2 \GeV & $ax+b$ && $2868.3\pm1.6$ & $2870.4\pm1.5$ & $2870.0\pm1.1$ & $2869.9\pm1.0$ \\\midrule
$e^+e^- \to \nu_\mu \bar{\nu}_\mu$ &$\lambda$&& $0.8370\pm0.0005$ & $0.8368\pm0.0003$ & $0.8368\pm0.0003$ & $0.8371\pm0.0003$ \\
500 \GeV & $ax+b$ && $0.8370\pm0.0004$ & $0.8368\pm0.0003$ & $0.8366\pm0.0003$ & $0.8373\pm0.0003$ \\\midrule
$e^+e^- \to ZH$ &$\lambda$&& $0.1209\pm0.0006$ & $0.1212\pm0.0006$ & $0.1220\pm0.0006$ & $0.1223\pm0.0006$ \\
365 \GeV & $ax+b$ && $0.1210\pm0.0007$ & $0.1211\pm0.0006$ & $0.1218\pm0.0007$ & $0.1219\pm0.0007$ \\
\bottomrule
\end{tabular}
\caption{Leading-order cross sections in the $G_\mu$ scheme using \Amegic.
$\epsilon$ is varied between $10^{-9}$ and $10^{-6}$, and $\delta=10^{-4}$ is used throughout.
The errors are statistical uncertainties on the Monte Carlo integration.
\label{table:methods:xs}}
\end{table}

\begin{table}
\centering
\begin{tabular}{lcrcccc}\toprule
\textbf{Process} & \textbf{Rescaling}&& \multicolumn{4}{c}{\textbf{Cross section (pb)}} \\
&&$\delta$ & $10^{-7}$&$10^{-6}$&$10^{-5}$&$10^{-4}$\\
\midrule
$e^+e^- \to \nu_\mu \bar{\nu}_\mu$ &$\lambda$&& $2876.0\pm7.0$ & $2870.7\pm3.3$ & $2872.0\pm1.8$ & $2872.1\pm1.2$ \\
91.2 \GeV & $ax+b$ && $2871.1\pm8.0$ & $2873.0\pm3.8$ & $2871.6\pm2.1$ & $2871.6\pm1.2$ \\\midrule
$e^+e^- \to \nu_\mu \bar{\nu}_\mu$ &$\lambda$&& $0.8393\pm0.0012$ & $0.8381\pm0.0007$ & $0.8376\pm0.0005$ & $0.8375\pm0.0003$ \\
500 \GeV & $ax+b$ && $0.8400\pm0.0014$ & $0.8384\pm0.0007$ & $0.8378\pm0.0005$ & $0.8376\pm0.0004$ \\\midrule
$e^+e^- \to ZH$ &$\lambda$&& $0.1236\pm0.0010$ & $0.1235\pm0.0005$ & $0.1230\pm0.0003$ & $0.1231\pm0.0002$ \\
365 \GeV & $ax+b$ && $0.1248\pm0.0012$ & $0.1236\pm0.0006$ & $0.1232\pm0.0003$ & $0.1230\pm0.0002$ \\
\bottomrule
\end{tabular}
\caption{Leading-order cross sections in the $G_\mu$ scheme using \Amegic.
$\delta$ is varied between $10^{-7}$ and $10^{-4}$, and $\epsilon=10^{-8}$ is used throughout.
The errors are statistical uncertainties on the Monte Carlo integration.
\label{table:methods:xsVaryDelta}}
\end{table}

\paragraph{Overestimating the ratio of structure functions.}

As a result of the formalism of the veto algorithm in
paragraph \ref{sec:methods:veto}, and the definition of
the initial-state splitting function in Eq.\ \eqref{eq:methods:sf_fratio},
in order to
use the veto algorithm to implement our parton shower we
require an analytically integrable overestimate of the splitting
probability, including the ratio of the initial-state parton
luminosities before and after the splitting.
Since the electron structure function $f_e(x)$ is singular as $x\to 1$,
this ratio can grow very large even when employing one of
the methods above to regularise it, and thus overestimating
it is non-trivial.
In this following we present a general method relying on
the asymptotic properties of the LL structure function.

Starting from the $x\to 1$ limit of the electron structure
function of eq.\ \eqref{eq:methods:electronSF_limit} the
luminosity ratio, including the factor $\tfrac{1}{z}$
present in the initial state splitting function,
goes as
\begin{equation}
  \frac{f_e\left(\tfrac{x}{z},t\right)}{z\,f_e(x,t)}
  \stackrel{x\to 1}{{}-\!\!\!-\!\!-\!\!\!\longrightarrow}
  \frac{1}{z}\left[\frac{1-x}{1-\frac{x}{z}}\right]^{1-\beta}
  <
  \frac{1}{z}\left[\frac{1-x}{1-\frac{x}{z}}\right]
  < \frac{1}{z-x}
  \,,
\end{equation}
where the inequalities hold since $0<x<1$ and $0<z<1$.
While $\beta=\beta(t)$ itself is unknown before $t$ is
chosen, $0<\beta<1$ for the whole kinematic range of all
colliders that can be envisioned currently.
Hence we get a worst-case singularity of $(z-x)^{-1}$
in the $x\to 1$ regime.
Extending it to the full $x$ range, including hard emission
corrections, we find that $k/(z-x)$ serves as an
overestimate with $k\approx 10$.

\paragraph*{The parton shower $\xmax$.}

With the above regularisation of the electron structure function,
the effect of a parton shower emission causing a transition between
$x<1-\delta$ and the $1-\delta<x<1$ region is under control.
This is aided by the behaviour of the shower upon approaching this
threshold. We choose to disallow any (further) evolution of an
initial-state lepton which has $x\geq \xmax =1-\delta$.
However, if a (potentially hard) emission would result in the
production of such a lepton, it is still allowed. This choice
captures the dominant contribution of the $[\xmax,1]$ phase space
region to the parton shower prediction and ignores only unresolvable
emissions within this region.

\subsection{The \texorpdfstring{\MCatNLO}{MC@NLO} method at NLO EW}
\label{sec:methods:mcatnlo}

The \MCatNLO method \cite{Frixione:2002ik},
in particular in the variant implemented in \Sherpa
\cite{Hoeche:2011fd,Hoeche:2012fm}, allows the automated
matching of an NLO QCD calculation to a parton shower
in a process-independent way.
With the QED parton shower of the previous section at hand,
we are now in a position to extend it to match NLO EW
corrections for electron-positron collider processes,
although purpose-built \NLOPS calculations for selected
processes, $e^+ e^- \to f\bar{f} (\gamma)$,
$e^+ e^- \to \gamma \gamma$, and more recently
$e^+ e^- \to \pi^+ \pi^- (\gamma)$, exist in
the event generator \BabaYaga \cite{Balossini:2006wc,Balossini:2008xr,Budassi:2024whw,Budassi:2026lmr}.
In this section, we describe the \MCatNLO method for
electroweak corrections as implemented in \Sherpa, and
highlight its differences with respect to its established
NLO QCD sister.
The methods described here allow the automated,
process-independent matching of any process at NLO EW to
a QED parton shower.\footnote{By working with an interleaved
QCD+QED parton shower, additive NLO QCD+EW
matching proceeds in the same way \cite{Flower:2026lui}.}

The core idea of the \MCatNLO method is to use the parton
shower to subtract all divergences at NLO.
To this end, in full anology to Eq.\ \eqref{eq:methods:PSfunc},
we introduce the modified parton shower functional
\begin{equation}\label{eq:methods:PSmodfunc}
  \begin{split}
    \bar{\mc{F}}_n(t_n,O)
    \,=\;&
      \bar{\Delta}_n(t_c,t_n) \, O(\Phi_n)
      +
      \sum_{\{\ajt,\bt\}} \sum_{\mr{f}}
      \int_{t_c}^{t_0}\done t\done z\frac{\done\phi}{2\pi}\;
      \overline{\mf{S}}_{a(b)\to \ajt \,\mr{f}\, (\bt)}(t,z,\phi)\,
      \bar{\Delta}_n(t,t_n) \,
      \mc{F}_{n+1}(t,O)\;.
  \end{split}
\end{equation}
Instead of the spin-averaged splitting functions, this modified shower uses
the full spin-dependent variants
$\mf{S}_{a(b)\to \ajt \,\mr{f}\, (\bt)}$,\footnote{
  As this paper is largely concerned with the intricacies of
  an initial state QED parton shower at lepton colliders and
  its matching to full NLO EW correction, we have chosen to
  use the initial-initial dipole notation throughout.
  Nonetheless, the algorithm is completely general and equally
  applies to all four types in depicted in
  Fig.\ \ref{fig:methods:sf:dips}.}
augmented with the corresponding initial-state luminosity ratio as in
Eq.\ \eqref{eq:methods:sf_fratio}.
This form captures the full infrared
divergence structure at NLO EW.
Likewise, its Sudakov factor is defined as
\begin{equation}\label{eq:methods:Sudmod}
  \begin{split}
    \bar{\Delta}_n(t',t_c)
    \,=\;&
      \exp\left[
        \sum_{\{\ajt,\bt\}} \sum_{\mr{f}}
        \int_{t_c}^{t_n}\hspace*{-5pt}\done t
        \int_{\zmin}^{\zmax}\hspace*{-10pt}\done z
        \int_0^{2\pi}\frac{\done\phi}{2\pi}\;
        \overline{\mf{S}}_{a(b)\to \ajt \,\mr{f}\, (\bt)}(t,z,\phi)\,
      \right].
  \end{split}
\end{equation}
Beyond NLO, if a spin-correlated emission has occurred, the
standard spin-averaged parton shower $\mc{F}$ continues the
evolution of the $(n+1)$--parton state.
Using the functionals defined in Eq.\ \eqref{eq:methods:PSfunc}
and Eq.\ \eqref{eq:methods:PSmodfunc},
we can thus write the expectation value for an
arbitrary infrared-safe observable $O$ as
\begin{equation}\label{eq:methods:mcatnlo}
  \begin{split}
    \langle O\rangle^\text{\MCatNLO}
    \,=\;&
      \intdphi{n}
      \bar{\mr{B}}(\Phi_n)\;\bar{\mc{F}}_n(t_n,O)
      +
      \intdphi{n+1}
      \mr{H}(\Phi_{n+1})\;\mc{F}_{n+1}(t_{n+1},O)\;,
  \end{split}
\end{equation}
where the $\mr{H}$ function to be defined below
contains contributions in the real-emission phase space,
while the $\bar{\mr{B}}$ function,
\begin{equation}\label{eq:methods:mcatnlo:Bbar}
  \begin{split}
    \bar{\mr{B}}(\Phi_n)
    \,=\;&
      \mr{B}(\Phi_n)
      +\tilde{\mr{V}}(\Phi_n)
      +\sum_{a,j,b}\int\done\Phi_1^{aj,b}\;
        \mr{D}_{aj,b}^\text{A}(\Phi_n\!\!\cdot\!\Phi_1^{aj,b})
      \,,
  \end{split}
\end{equation}
collects all contributions which occur in the Born phase space.
Besides the Born matrix element $\mr{B}$ itself, these are
the renormalised virtual correction $\tilde{\mr{V}}$,
defined to include the collinear counterterms, and the parton shower
splitting kernels integrated over the entire soft-collinear
phase space up to the $n$-parton shower starting scale $t_n$,
\begin{equation} \label{eq:methods:mcatnlo-corr}
  \begin{split}
    \int\done\Phi_1^{aj,b}\;
    \mr{D}_{aj,b}^\text{A}(\Phi_n\!\!\cdot\!\Phi_1^{aj,b})
    \,=\;&
      \mr{B}(\Phi_n)\otimes
      \int_{0}^{t_n}\hspace*{-5pt}\done t
      \done z\frac{\done\phi}{2\pi}\;
      \overline{\mf{S}}_{a(b)\to \ajt \,j\, (\bt)}(t,z,\phi)
      \;.
  \end{split}
\end{equation}
By comparison with Eq.\ \eqref{eq:methods:SF:approx-corr},
it is evident that the infrared poles of Eq.\ \eqref{eq:methods:mcatnlo-corr}
are identical to those
of the full real-emission matrix element and, thus,
cancel those of the virtual correction by virtue of the
KLN theorem, rendering the $\bar{\mr{B}}$ function
IR finite.
In general, however, the coefficients of the Laurent
expansion of the integrated parton shower splitting kernels
in $D$ dimensions are unknown. Hence, we employ a second subtraction,
\begin{equation}\label{eq:methods:mcatnlo:double-sub}
  \begin{split}
    \sum_{a,j,b}\int\done\Phi_1^{aj,b}\;
    \mr{D}_{aj,b}^\text{A}(\Phi_n\!\!\cdot\!\Phi_1^{aj,b})
    \,=\;&
      \sum_{\ajt,\bt}\mr{I}_{\ajt,\bt}^\text{S}(\Phi_n)
      +\sum_{a,j,b}\int\done\Phi_1^{aj,b}
        \left[
          \mr{D}_{aj,b}^\text{A}(\Phi_n\!\!\cdot\!\Phi_1^{aj,b})
          -\mr{D}_{aj,b}^\text{S}(\Phi_n\!\!\cdot\!\Phi_1^{aj,b})
        \right]
      \,,\hspace*{-20mm}
  \end{split}
\end{equation}
using the real and integrated subtraction terms,
$\mr{D}_{aj,b}^\text{S}$ and $\mr{I}_{\ajt,\kt}^\text{S}$,
of the Catani-Seymour subtraction formalism
\cite{Catani:1996vz,Catani:2002hc,Schonherr:2017qcj}.
This allows the simple evaluation of the left-hand side of
Eq.\ \eqref{eq:methods:mcatnlo:double-sub} when the choice
\begin{equation} \label{eq:methods:smcatnlo}
  \mr{D}^\mr{A}_{aj,b} = \mr{D}^\mr{S}_{aj,b} \Theta(\mu_Q^2-t_{aj,b})
\end{equation}
is made for the parton shower splitting kernels
\cite{Hoeche:2011fd}.

Turning again to the $\mr{H}$ function in Eq.\ \eqref{eq:methods:mcatnlo},
we now specify its definition as
\begin{equation}\label{eq:methods:mcatnlo:H}
  \begin{split}
    \mr{H}(\Phi_{n+1})
    \,=\;&
      \mr{R}(\Phi_{n+1})
      -\sum_{a,j,b}\mr{D}_{aj,b}^\text{A}(\Phi_n^{aj,b}\!\!\cdot\!\Phi_1^{aj,b})
      \;,
  \end{split}
\end{equation}
with the real-emission contribution $\mr{R}$ being
subtracted by the soft-collinear approximation
generated by our parton shower.
$\Phi_{n+1}=\Phi_n^{aj,b}\!\!\cdot\!\Phi_1^{aj,b}$
is the dipole-dependent factorised phase space.
With these definitions for $\bar{\mr{B}}$
and $\mr{H}$ as well as our modified parton shower,
it is straightforward to show that the $\order(\alpha)$
expansion of Eq.\ \eqref{eq:methods:mcatnlo}
correctly reproduces the NLO expression for any
infrared-safe observable $O$ \cite{Hoeche:2011fd}.

Further, it is important to note from Eq.\ \eqref{eq:methods:smcatnlo}
the phase space limit in $\mr{D^\text{A}}$,
namely that it is strictly zero
for emissions with $t>t_n$.
Hence, while soft emissions will be described through
our (modified) parton shower resummation in
$\bar{\mc{F}}_n$ acting on $\bar{\mr{B}}$,
corrected to full $\order(\alpha)$ accuracy by $\mr{H}$,
hard emissions are solely described by $\mr{H}$.

\paragraph*{The \MCatNLO algorithm.}
The \MCatNLO method according to Eq.\ \eqref{eq:methods:mcatnlo}
is implemented using the following algorithm.
\begin{enumerate}
  \item
    Generate one of two types of events:
    either an $\mb{S}$-event (`soft' or `standard') using
    the $\bar{\mr{B}}$ function of Eq.\ \eqref{eq:methods:mcatnlo:Bbar},
    or an $\mb{H}$-event (`hard') using the $\mr{H}$ function
    of Eq.\ \eqref{eq:methods:mcatnlo:H}.
    If unweighted events are sought, their relative
    composition should be that of their respective
    cross sections.
  \item
    If an $\mb{S}$-event was chosen, feed the resulting Born-type
    phase space configuration to the fully spin-correlated
    QED parton shower $\bar{\mc{F}}_n$ of Eq.\ \eqref{eq:methods:PSmodfunc},
    using the hard scale $t_n$ as its starting scale.
    If an emission is generated, pass the event
    to the standard shower $\mc{F}_{n+1}$ which
    continues the evolution from the newly-generated scale
    $t_{n+1}$.
  \item
    If an $\mb{H}$-event was chosen, pass its
    real-emission-type phase-space configuration directly
    to the standard parton shower $\mc{F}_{n+1}$.
    Determine its starting scale $t_{n+1}$ by interpreting
    its existing emission kinematics as a parton splitting
    process, using the inverse of $\mc{F}_n$ as a clustering algorithm \cite{Hoeche:2009rj}.
\end{enumerate}

\subsection{Scale choices}
\label{sec:methods:scales}

The choice of factorisation scale in an \MCatNLO EW
is relevant despite the slow running of $\alpha$ compared
to that of $\alpha_s$.
As has been argued \cite{Hoeche:2011fd,Nagy:2003tz},
a lack of phase-space restrictions on the exponentiated
($\overline{\mf{S}}$ and $\overline{\mr{S}}$)
part of the real emission can lead to logarithmic contributions to
\MCatNLO of $\alpha \log^2{(q^2/s)}$, where $q^2$ is the virtuality of the
emitter parton, instead of the correct parton shower contribution of
$\alpha \log^2{(q^2/\mu_F^2)}$.
While the leading logarithms are unaffected by this starting
scale choice, higher-order logarithms differ.
This implies that where $\mu_F^2 \ll s$, a careful choice of scale is needed.
The scale chosen must be soft- and collinear-safe, so the invariant mass of
bare charged particles cannot be used to define the scale.
In the case of the processes studied here, we use the invariant mass of the
uncharged final state as the factorisation scale.

Similarly, the shower starting scale $\mu_Q^2$ should be a typical scale of
the hard process.
The purpose of this scale is twofold: to put an upper limit on the kinematics
of emissions generated using splitting functions, and to ensure that the correct
Sudakov factors are applied to all emissions.
For these reasons, the starting scale is treated slightly differently between
$\mathbb{S}$- and $\mathbb{H}$-events, in order to reflect the different physical
descriptions of radiation that these imply.

In an $\mathbb{S}$-event, the ordinary shower must be a continuation of the
\MCatNLO shower, even if the splitting functions are slightly different.
This results in an identification of the factorisation scale $\mu_F$ which
constrains the contributions of the soft exponentiated terms $\mr{D^A}$ to be the
same as the $\mathbb{S}$-event \MCatNLO shower starting scale $\mu_Q$.
Hence, the \MCatNLO shower starts from $t_n=\mu_Q^2$ and either
continues to an IR cutoff (in which case the shower cannot act and there must
be no further emissions), or generates an emission at $t_1$.
In both cases a Sudakov factor is generated.
The ordinary shower then continues from $t_{n+1}=t_1$ to the infrared cutoff,
which must be the same as the \MCatNLO cutoff for consistency.

For the $\mathbb{H}$-events, to preserve the total NLO cross section, there can be no Sudakov factor applied to the first emission as there is no splitting function term to reinstate unitarity\footnote{In NLO merging, where the total cross section will in any case not be the NLO cross section, applying the Sudakov here is perfectly acceptable.}.
However, to avoid double counting, this emission must be reinterpreted in the parton shower language to have scale $t_{n+1}$.
This is achieved using the \MEPS clustering algorithm \cite{Hoche:2010kg}.
Now we consider the two possible cases:

\hspace{20pt}$\mathbf{t_{n+1}<\mu_Q^2}$\hspace{10pt} The shower should start from $t_{n+1}$,
since to allow a harder emission would amount to a rejection of the matching and
resummation principles by allowing an unordered configuration.

\hspace{20pt}$\mathbf{t_{n+1}>\mu_Q^2}$\hspace{10pt} This case is not clear-cut from
resummation arguments. On the one hand, $t_{n+1}$ is now the hardest scale of the process,
but on the other, we would like to avoid extending the resummation beyond its region
of validity.
Here we choose to start the shower from $t_{n+1}$, which extends both the emission
probability and the Sudakov factor up to the new hardest scale.

\paragraph*{Photons at Born level.}
The presence of photons in the Born matrix element complicates the implementation
of the \MCatNLO method slightly.
Clearly, the splitting of such a photon is of the same perturbative order as the
emission of a photon off any charged particle in the process; however, analyses
frequently use isolated or tagged hard `prompt' photons.
As such, events in which a Born photon splits in an $\mathbb{S}$-event will be rejected
by the analysis criteria unless a similar photon is emitted by the parton shower.
In practice, one can label a photon from the matrix element as
tagged, and ensure that it becomes a long-distance photon (and
couples with $\alpha(0)$). However, if the Born matrix element
contains two (or more) photons, only one of which is tagged while the other may
split, further consideration is needed. If photon splittings do
occur, care must be taken to match their spectator schemes
between the $\mathbb{S}$- and $\mathbb{H}$-events.

\subsection{\texorpdfstring{$\alpha$}{Alpha} and electroweak input schemes} \label{sec:methods:ewscheme}

Before presenting results from the QED parton shower and \MCatNLO,
we briefly discuss EW input schemes and the treatment of $\alpha$.
For the evaluation of MEs in hard processes, it is most appropriate
to use a value of $\alpha$ which resums higher-order corrections.
Using $\alpha$(0) for processes at scales of the EW gauge bosons
or higher leads to logarithms of the type $\log (m^2_\mr{f} /m^2_{Z/W} )$,
which are large for light flavours $\mr{f}$.
To include these photonic vacuum polarisation corrections to the
coupling, a short-distance renormalisation scheme must be used,
commonly either of the $\alpha(m_Z)$ or $\Gmu$ schemes.
In particular, the $\Gmu$ scheme absorbs higher-order corrections
to the renormalisation of the weak mixing angle, and thus is often
the scheme of choice when a fermion interacts with a weak boson
at Born level.

Since the difference between the values of $\alpha$ in different
schemes is formally subleading, we have the freedom to define a
different scheme for radiative corrections \cite{Denner:2019vbn}.
In the case of NLO calculations, of course, this must be consistent
in all parts of the NLO calculation to ensure cancellation of
singularities.
A similar procedure was followed, for example, in refs.~\cite{Greiner:2017mft,Gutschow:2020cug,Pagani:2021iwa},
to consistently use the correct $\alpha$ in different parts of an
NLO calculation.
For the case of the shower, which is unitary, we are free to use a
running $\alpha$ in the shower, analogously to QCD.
For photon emissions, since most will become long-distance photons
(i.e.\ they will not split again), their QED coupling should be evaluated
in the Thomson limit, so we choose $\alpha(0)$, capturing
the photon-to-photon fragmentation function to $\order(\alpha)$.
This is because the photonic wavefunction renormalisation already
exactly cancels the light-fermion logarithms in the renormalisation
of the coupling constant.
For photon splittings into fermions, we choose the coupling evolved
to the splitting scale, $\alpha(t)$, where here $t$ is the reduced
virtuality.
Further, if the splitting photon was produced with its coupling
set to $\alpha(0)$, we additionally include a correction to its
coupling factor of $\alpha(\bar{q}^2)/\alpha(0)$ in its
splitting function, where $\bar{q}^2$ is the virtuality of
the splitting photon of
Eq.\ \eqref{eq:methods:IIkTordering}-\eqref{eq:methods:FFkTordering}.
Similarly to the CMW scheme in QCD \cite{Catani:1990rr} which
recovers the soft cusp anomalous dimension from the running
of the strong coupling in the \MSbar scheme, this scale choice
recovers higher-order logarithms associated with the $\beta$-function
from the running QED coupling in the on-shell scheme
\cite{Brodsky:1982gc}.
Additionally, in the electron structure function, we use $\alpha(0)$.
This is consistent with physical expectations and with existing
conventions.

\section{Validation: \texorpdfstring{$e^+ e^- \to \nu_\mu \bar{\nu}_\mu$}{ee->nunu}}
\label{sec:validation}

Having outlined the necessary modifications to the \MCatNLO
method for use with a QED parton shower, it now remains to
test the implementation and stability of the method.
We will examine the dependence of our predictions on the
parameters $\epsilon$, $\delta=1-\xmax$, and the infrared
cutoff $t_c$.
In addition to demonstrating the numerical stability of the
method, this will inform the choice of values for these
parameters to produce predictions for $e^+ e^- \to ZH$
in Sec.\ \ref{sec:results}.
After validating the QED parton shower this way, we will test
that the \MCatNLO fully spin-correlated emission reproduces
its results upon neglecting the hard-emission corrections
which are needed to make the \MCatNLO prediction correct to
NLO.
Finally, in the next section
we will compare the LO+PS and \MCatNLO predictions
to the fixed-order LO and NLO for $ZH$ production.

In the present section, we will set up simple test scenarios
to examine a range of observables to validate the QED parton
shower and fully spin-correlated \MCatNLO.
Throughout, we use the matrix element generators \Amegic
\cite{Krauss:2001iv}, \Comix \cite{Gleisberg:2008fv} and
\OpenLoops \cite{Buccioni:2019sur,Denner:2016kdg,
  Ossola:2007ax,vanHameren:2010cp} for the tree-level
matrix elements, renormalised one-loop amplitudes, and
infrared subtraction.
We use the LL electron structure function \cite{Kuraev:1985hb},
see Sec.\ \ref{sec:methods:electronSF}, with the electron
mass parameter $m_e=511\,\text{keV}$ and the dynamic
scale $Q^2=s'=m_{\nu\bar{\nu}}^2$ in the hard scattering process.
We work in the complex mass scheme \cite{Denner:2005fg},
\begin{equation}
  \begin{split}
    \mu_i^2 = m_i^2-\mr{i}m_i\Gamma_i
    \qquad\qquad
    \text{with}
    \qquad\qquad
    i\in\{W,Z,H,t\}\,,
  \end{split}
\end{equation}
with the following masses and widths
\begin{equation}
  \begin{array}{rclcrcll}
    m_W      & \!=\! & \phantom{0}80.370\phantom{0}\,\text{GeV} & \qquad
    \Gamma_W & \!=\! & 2.0897\,\text{GeV} \\
    m_Z      & \!=\! & \phantom{0}91.1876\,\text{GeV} & \qquad
    \Gamma_Z & \!=\! & 2.4952\,\text{GeV} \\
    m_H      & \!=\! & 125.09\phantom{00}\,\text{GeV} & \qquad
    \Gamma_H & \!=\! & 0.0041\,\text{GeV} \\
    m_t      & \!=\! & 172.5\phantom{000}\,\text{GeV} & \qquad
    \Gamma_t & \!=\! & 1.32\phantom{00}\,\text{GeV}&.
  \end{array}
\end{equation}
All other particles are treated massless in the
perturbative matrix elements.
In the parton shower, external particles are shifted onto
their physical mass shells and evolve according to their
massive splitting functions, detailed in App.\ \ref{app:sfs}.

As described in Sec.\ \ref{sec:methods:ewscheme}, we use the
$G_\mu$ scheme with $(\Gmu,\mu_W^2,\mu_Z^2)$ as our input
parameter set to define and renormalise our EW parameters.
The Fermi constant itself takes the value
\begin{equation}
  \Gmu = 1.16639\times 10^{-5}\,\text{GeV}^{-2}\;.
\end{equation}
Consequently, it defines the lowest-order couplings.
To match the NLO corrections to the parton shower, however,
we use the $\alpha(0)$ scheme with $(\alpha(0),\mu_W^2,\mu_Z^2)$
for the additional coupling entering at NLO and in the photon
emissions in the parton shower.
Their values are given by
\begin{equation}
  \begin{split}
    \alpha_{\Gmu}
    \,=\;
      \left|\frac{\sqrt{2}\,s_\text{w}^2\,\mu_W^2\,G_\mu}{\pi}\right|
    \qquad\qquad
    \text{and}
    \qquad\qquad
    \alpha(0)
    \,=\;
      1/137.03599976\;,
  \end{split}
\end{equation}
with the weak mixing angle
\begin{equation}
  \begin{split}
    s_\text{w}^2=1-c_\text{w}^2=1-\frac{\mu_W^2}{\mu_Z^2}\;.
  \end{split}
\end{equation}
For the study of the dependence of
initial-state radiation on the shower parameters, and for
easier interpretation of the results,
we consider only photon radiation off the initial leptons
in this and the following section.
However, photon splittings into leptons and quarks
are fully implemented in the parton shower
as described in App.\ \ref{app:sfs},
and the relevant switches are detailed in App.\ \ref{app:settings}.
The effect of producing further charged-particle pairs can
be very relevant depending on the observable \cite{Flower:2022iew},
but here we focus on the radiative impacts on the initial-state
particles, which translate to recoil of the entire final state;
where the initial-state particles are leptons, these impacts
are dominated by their photon emissions.

In order to validate the implementation of the QED parton shower
for initial-state leptons, we study muon neutrino production at
an electron-positron collider for two different centre-of-mass
energies: $\sqrt{s}=91.2\,\text{GeV}$ and $\sqrt{s}=500\,\text{GeV}$.
While the former is dominated by soft radiation, the latter
affords plenty of phase space and matrix element support to
collinear emissions.
The combination of both setups will allow us to examine the
shower's description of radiation in both limits.
The choice of muon neutrinos as the final state ensures that
only the $s$-channel $Z$ exchange diagram contributes, which
results in a clean resonance structure of the cross section
and a straightforward identification of the relevant scales,
see Sec.\ \ref{sec:methods:scales}.

In addition to examining specific radiation-induced observables,
we will study two which appear at LO:
the neutrino-pair invariant mass $m_{\nu\bar{\nu}}$, and the
(matter) neutrino transverse momentum $k_\mr{T}^\nu$.
The former is completely unchanged by initial-state showering
due to the specific kinematic construction of the Catani-Seymour
shower, so provides a useful cross-check of the implementation 
and allows comparison
of the LO and NLO distributions without shower effects.
The latter will provide an analogue to identified-final-state
transverse momenta, such as the $Z$ transverse momentum
which we will study in Sec.\ \ref{sec:results}, and is, in general,
modified from its fixed-order distribution by the action of a
parton shower.

The radiative observables that will be presented here include
the hardest photon transverse momentum
$k_\mr{T}^\gamma = \text{max}(k_\mr{T}^{\gamma_i})$,
the second-hardest photon transverse momentum $k_\mr{T}^{\gamma_2}$,
and the photon multiplicity $N_\gamma$ with transverse momentum
above a cut $k_\mr{T}^{\gamma_i} > k_\mr{T}^\text{min}$.
For this process, we choose $k_\mr{T}^\text{min}=0.5$\,GeV.
In addition, we examine the 0-to-1 QED jet rate $d_{01}$,
as defined using the $k_\mr{T}$ clustering algorithm.
In this algorithm, the distance parameter is given by 
\begin{equation}
  d_{ij}
  = \min(k_{\mr{T}i}^2,k_{\mr{T}j}^2) \,
    \frac{\Delta R_{ij}^2}{R^2},
\end{equation}
where $\Delta R_{ij}$ is the distance in pseudorapidity and
azimuthal angle,
$\Delta R_{ij}^2 = \Delta \eta_{ij}^2 + \Delta \phi_{ij}^2$.
We use a radius parameter $R=1$.
In this process, the particles which we consider as input to
the algorithm are electrons and photons.
The differential jet rate $d_{n,n+1}$ measures at what scale
the $(n+1)^\text{th}$ jet is formed, and is thus ideally suited
to assess the internal dynamics of our initial state evolution.
Where only photon emissions are modelled, this is equivalent
to a photon-$k_\mr{T}$-type observable (up to differences in
the shower evolution variable), but where photon splittings are
considered, the two will differ.
Using a jet algorithm to define a final state means that we are
inclusive with respect to flavour, analogously to QCD parton showers.

There remain some choices to be made in the shower algorithm and
in the analysis, which will be informed by the results of this
test process.
In particular, we study the various infrared cutoff parameters,
then validate the first emission in \MCatNLO as a special case
of the parton shower.

\subsection{Variation of infrared parameters}

First, we investigate the effects of the shower infrared cutoff,
$t_c$, on the observables described above, then we study the two
structure function parameters $\epsilon$ and $\delta$ defined in
Sec.\ \ref{sec:methods:electronSF}.
To do so, we restrict ourselves to the shower acting on LO events.
The NLO calculation, being afflicted with an additional integration
over the same integrable divergence in the electron structure
function, necessitates the introduction of another technical
parameter, $\varepsilon$, which is detailed in App.\ \ref{app:kp}.
We have checked that the dependence of the cross section on
its value is far smaller than that for the LO parameters
investigated here.

We have seen, in Tab.\ \ref{table:methods:xs} and
\ref{table:methods:xsVaryDelta}, that the variation of $\epsilon$
and $\delta$ within reasonable limits does not alter the total LO
cross section.
Additionally, by construction, the variation of $t_c$ does not
affect the total cross section, nor any Born-level observable with
a resolution larger than $t_c$.
However, the impact of the choice of these three variables on
differential observables must be evaluated, as well as their effect
on the efficiency of event generation.

\subsubsection{The shower infrared cutoff \texorpdfstring{$t_c$}{tc}}

\begin{figure}
  \centering
  \includegraphics[width=0.45\textwidth]{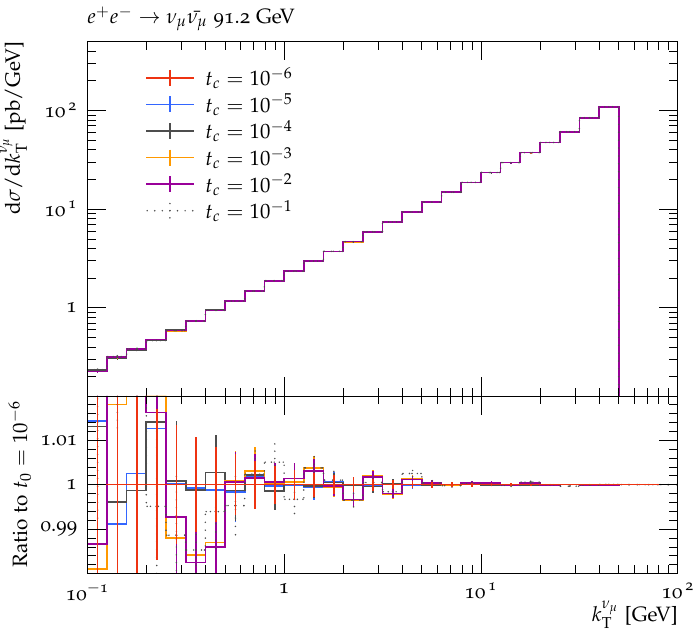}
  \includegraphics[width=0.45\textwidth]{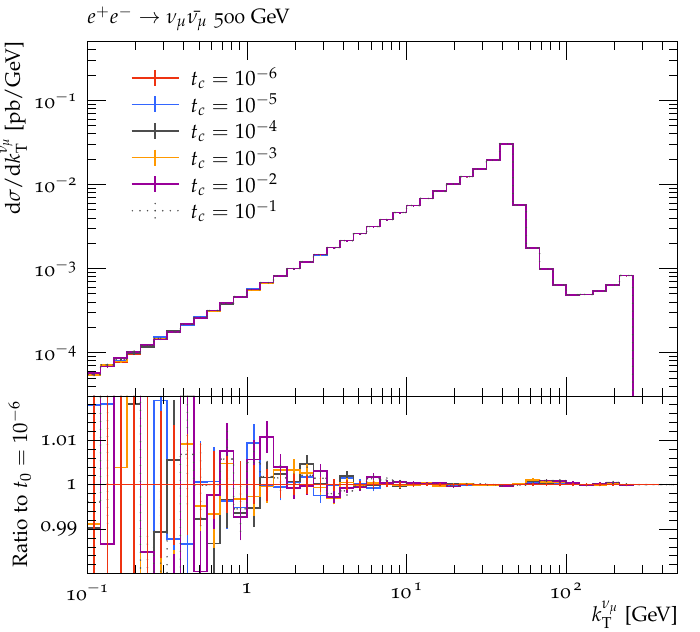}
  \hfill
  \includegraphics[width=0.45\textwidth]{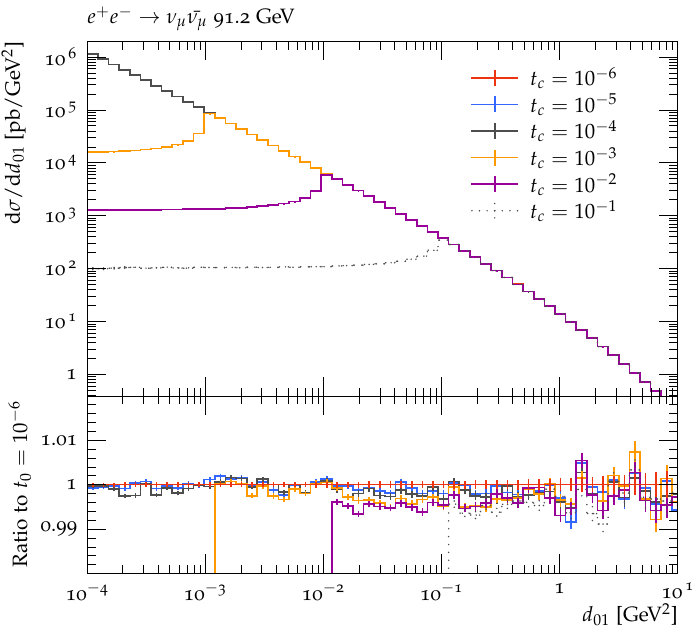}
  \includegraphics[width=0.45\textwidth]{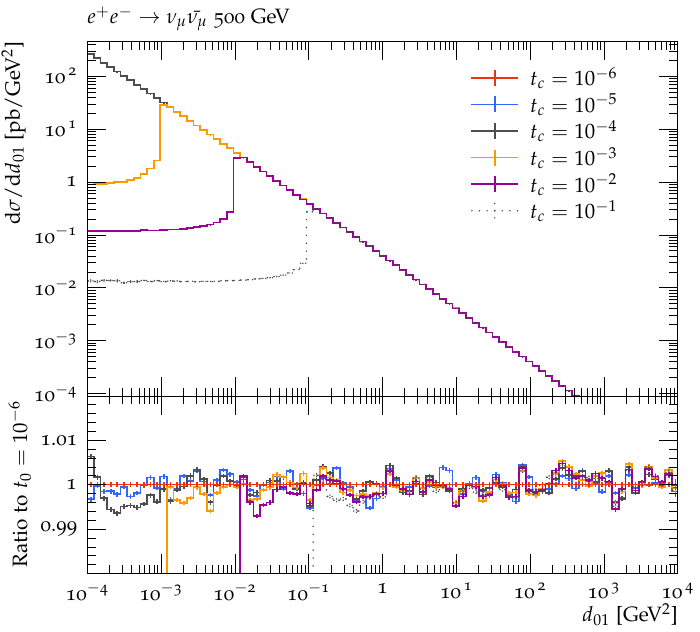}
  \hfill
  \includegraphics[width=0.45\textwidth]{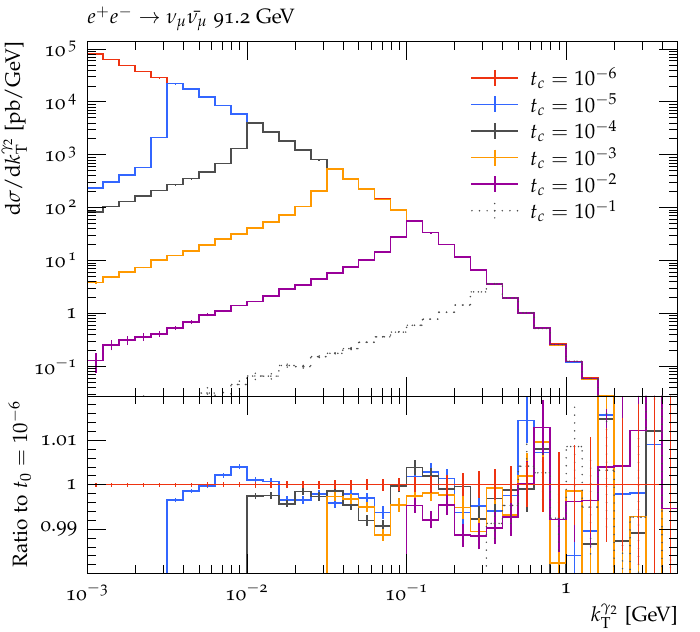}
  \includegraphics[width=0.45\textwidth]{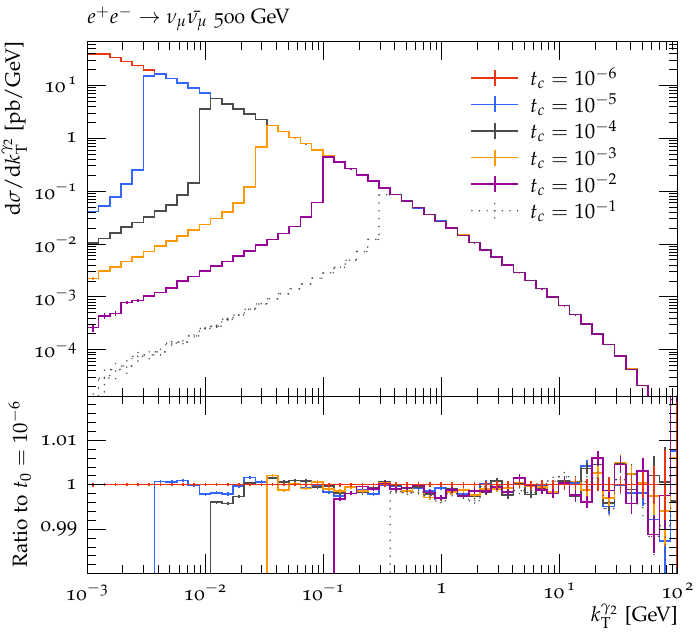}
  \caption{
    Plots showing the effect of varying the parton shower infrared cutoff,
    $t_c$, for the process $e^+ e^- \to \nu_\mu \bar{\nu}_\mu$ at two
    different collider energies, 91.2\,GeV (left) and 500\,GeV (right).
    The observables shown are the neutrino transverse momentum (top row),
    the 0-to-1 jet rate (middle) and the second photon transverse momentum
    (bottom).
    The infrared cutoff is varied logarithmically between $10^{-6}\,\GeV^2$
    and $10^{-1}\,\GeV^2$.
    \label{fig:results:varyt0}
  }
\end{figure}

In Fig.\ \ref{fig:results:varyt0}, we see that varying the shower
infrared cutoff, $t_c$ results in very little change to inclusive
Born-level observables such as the neutrino transverse momentum,
but a sharp cutoff is visible in radiative observables.
This is as expected, since the strict ordering and unitary construction
of the photon-emission-only shower ensure that an emission at a given
$t$ does not affect emissions which are harder in $t$.
The distributions in the 0-to-1 jet rate and second photon transverse
momentum do not drop to zero below the cutoff, however, since neither
variable corresponds exactly to the parton shower ordering variable.
It is important to note that although the differential cross sections
below the cutoff are not small (indeed, the underflow bin contributes
most of the cross section), the kinematic impact on the rest of the
event is very small for these soft photons, as illustrated by the
stability of the neutrino transverse momentum.
Thus, the choice of whether to produce them with the parton shower is
dependent on the desired observable resolution, informed by experimental
constraints such as detector limits.

The agreement in the radiative observables above the respective infrared cutoff is better
than 1\% for both collider energies, while the agreement in the neutrino transverse
momentum is on the 0.1\% level across the majority of phase space.
For the remainder of this section, we will use $t_c=10^{-6}\,\GeV^2$, to
avoid contaminating the soft end of the spectrum when varying the other
parameters.
However, because of the straightforward dependence of the radiative
observables on $t_c$, it can be varied anywhere below the sensitivity of
the observable in question without introducing further uncertainties.

\subsubsection{The \texorpdfstring{$\epsilon$}{epsilon} structure function parameter}

\begin{figure}
  \centering
  \includegraphics[width=0.45\textwidth]{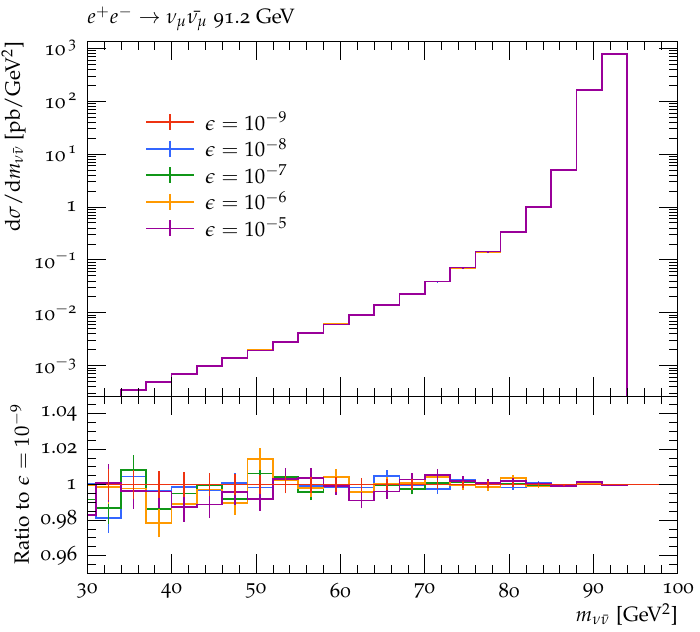}
  \includegraphics[width=0.45\textwidth]{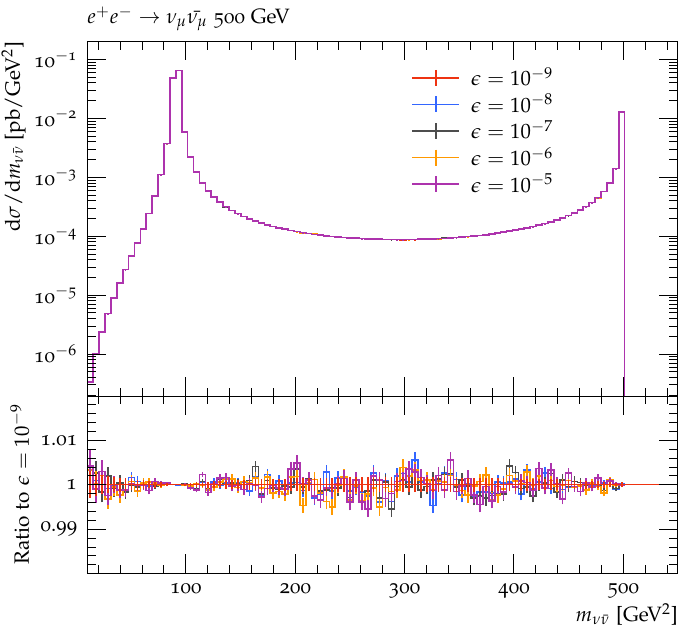}
  \hfill
  \includegraphics[width=0.45\textwidth]{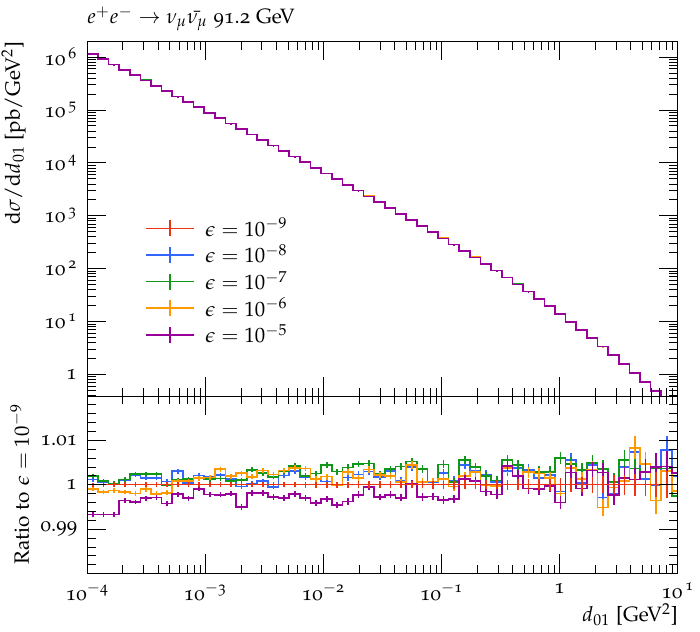}
  \includegraphics[width=0.45\textwidth]{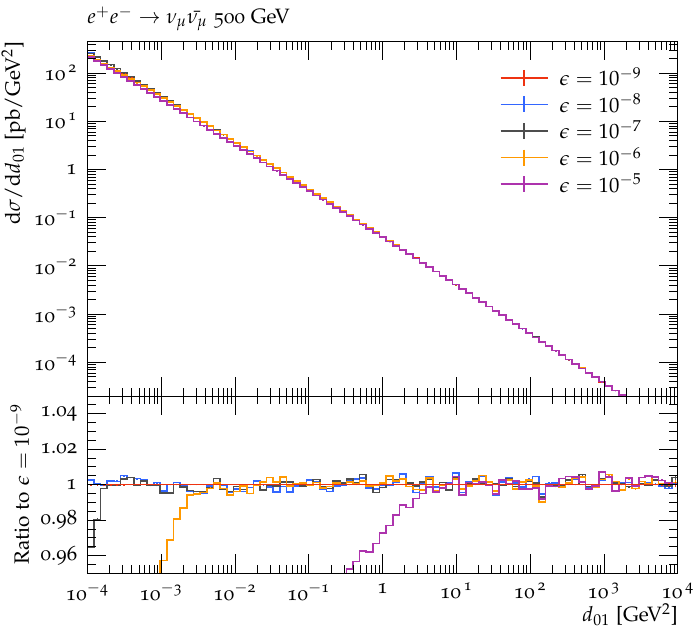}
  \hfill
  \includegraphics[width=0.45\textwidth]{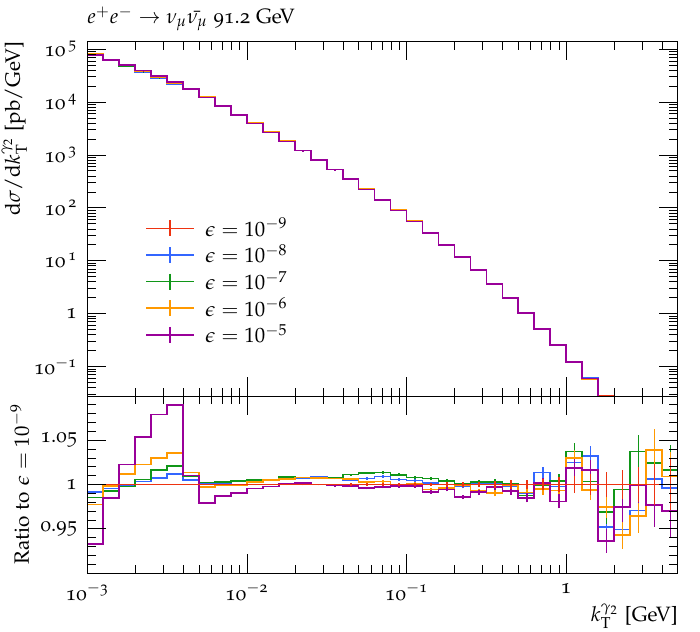}
  \includegraphics[width=0.45\textwidth]{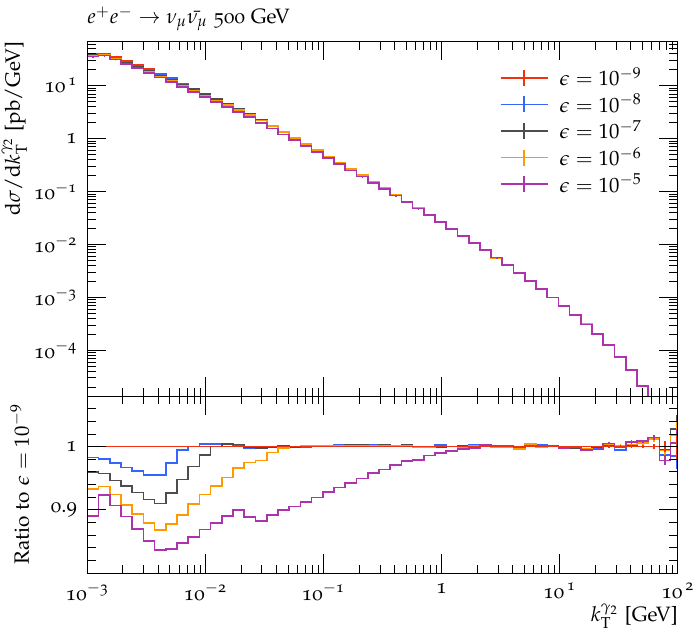}
  \caption{
    Plots showing the effect of varying the structure function
    $\epsilon$ parameter for the process
    $e^+ e^- \to \nu_\mu \bar{\nu}_\mu$ at two different collider
    energies, 91.2\,GeV (left) and 500\,GeV (right).
    The observables shown are the neutrino-pair invariant mass (top row),
    the 0-to-1 jet rate (middle) and the second photon transverse momentum
    (bottom).
    $\epsilon$ is varied logarithmically between $10^{-9}$ and $10^{-5}$.
    \label{fig:results:varyEps}
  }
\end{figure}

The top row of Fig.\ \ref{fig:results:varyEps} shows that varying
the structure function parameter $\epsilon$ does not affect Born-level
observables such as the neutrino-pair invariant mass at the sub-percent
level.
The invariant mass is a useful observable to compare, unlike for the
variation of $t_c$, because its dependence on $\epsilon$ enters
both before the parton shower evolution and during it, due to the ratio
of structure functions which appears.
The parton shower evolution will never move the event in phase space up
to the boundary $x=1-\epsilon$, but the value of $\epsilon$ affects the
rescaling near the boundary, which can change the parton shower splitting
probabilities.

Meanwhile, in the middle row we verify that varying $\epsilon$ between
$10^{-9}$ and $10^{-7}$ does not affect the radiative distribution in the
0-to-1 jet rate, for both collider energies.
In fact, we observe agreement at the sub-percent level,
showing the self-consistency of our linear rescaling method.
However, we see a systematic 0.5\% underestimate of the jet rate for the
largest $\epsilon$ value for $\sqrt{s}=91.2\,\text{GeV}$, and a much more
severe disagreement for $\epsilon = 10^{-5}$ and $10^{-6}$ at 500\,GeV.

Similarly, in the bottom row, we show the transverse momentum of the
second-hardest identified photon.
It is clear that this observable is more sensitive to the structure
function parameters, showing variation of a few percent for many values
of $\epsilon$.
Although larger, the deviations in $k_\mr{T}^{\gamma_2}$ are consistent
with those in $d_{01}$, and show that this method of rescaling the
structure function is only fully stable, for $\epsilon=10^{-8}$ compared
to $10^{-9}$, for photon emissions at a scale harder than $10^{-2}$\,GeV
for a collider at the $10^2$\,GeV scale.
Photons at this scale will have almost no kinematic impact and the photon
itself is very unlikely to be detectable.
For the application of this method to collisions at much lower energies,
the range of stable parameters will again have to be determined.

It is clear from Fig.\ \ref{fig:results:varyEps} that at the $Z$ mass,
$\epsilon=10^{-6}$ or smaller is sufficient to well-describe all studied
observables, while at 500\,GeV, the requirement is more stringent for
radiative observables.
For this reason, for the rest of this paper, we will use
$\epsilon=10^{-8}$.

\subsubsection{The \texorpdfstring{$\delta$}{delta} parameter/parton shower \texorpdfstring{$x_\text{max}$}{xmax}}

The $\delta$ parameter here controls the lower limit of the structure
function rescaling, as well as the upper limit on $x$ in the parton
shower, see Sec.\ \ref{sec:methods:electronSF}.
As a result, the aim of this validation is to find the largest value of
$\delta$ which makes reliable predictions, since this will allow faster
shower evolution.

\begin{figure}
  \centering
  \includegraphics[width=0.45\textwidth]{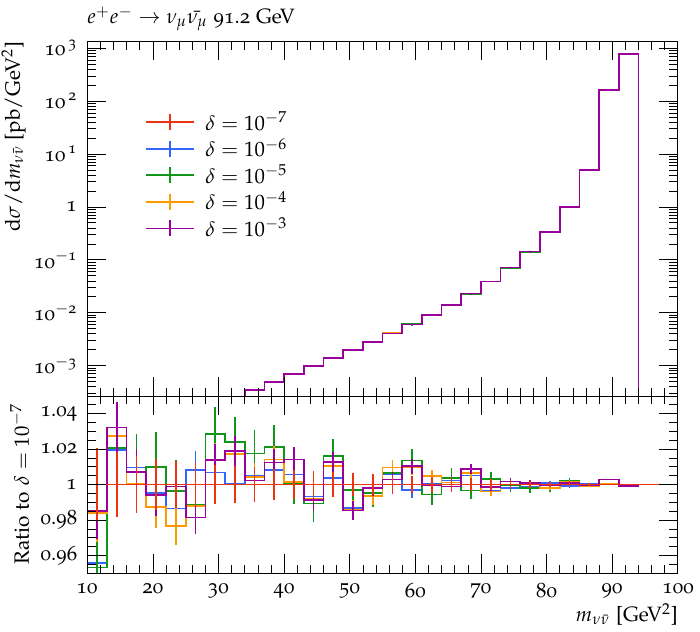}
  \includegraphics[width=0.45\textwidth]{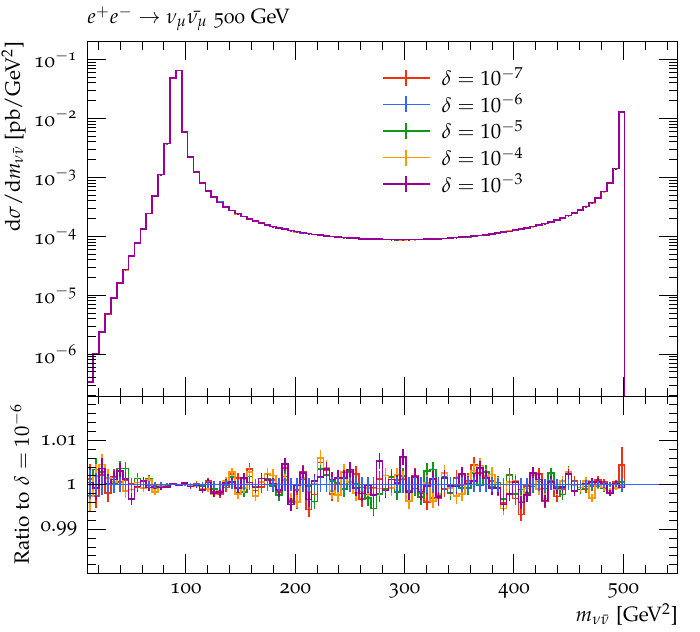}
  \hfill
  \includegraphics[width=0.45\textwidth]{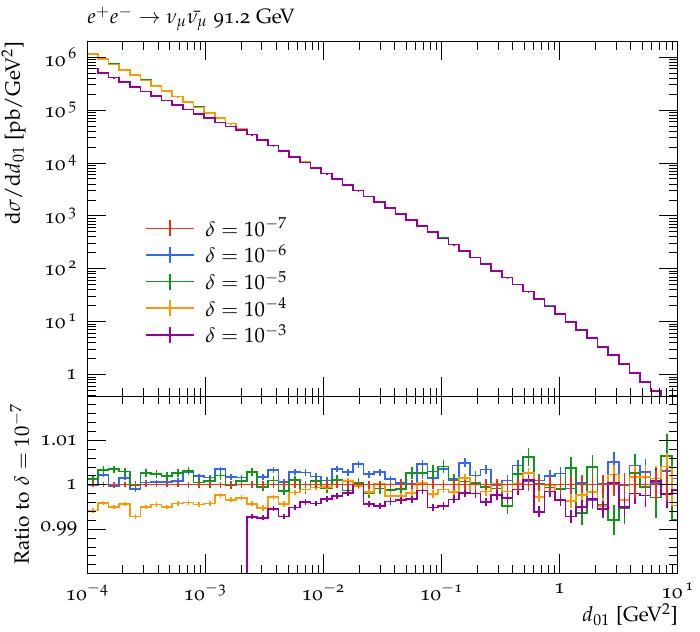}
  \includegraphics[width=0.45\textwidth]{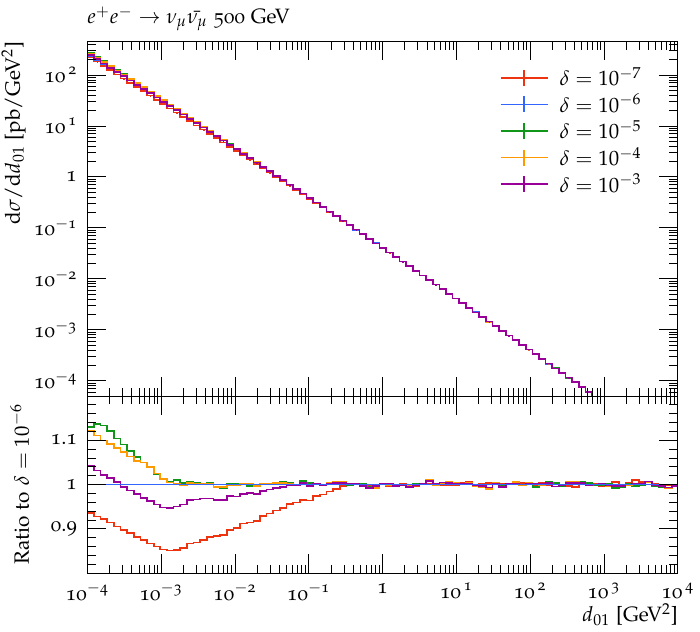}
  \hfill
  \includegraphics[width=0.45\textwidth]{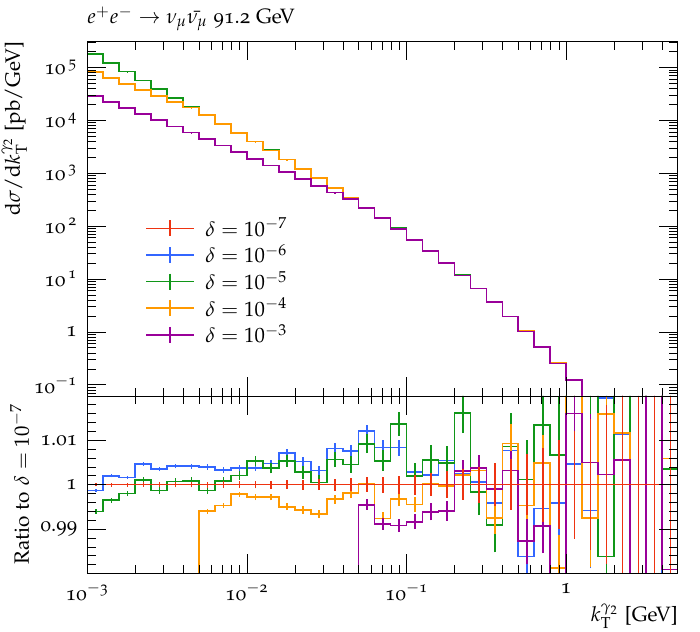}
  \includegraphics[width=0.45\textwidth]{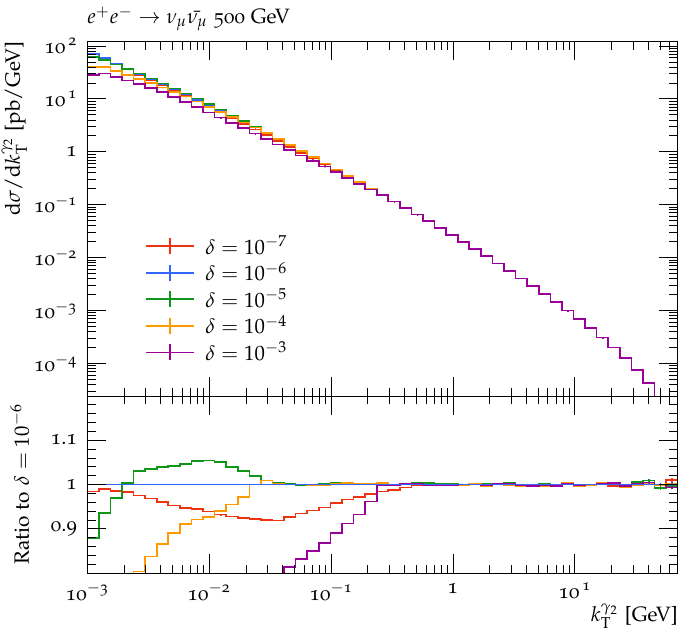}
  \caption{
    Plots showing the effect of varying the structure function $\delta$
    parameter for the process $e^+ e^- \to \nu_\mu \bar{\nu}_\mu$ at two
    different collider energies, 91.2\,GeV (left) and 500\,GeV (right).
    The observables shown are the neutrino-pair invariant mass (top row),
    the 0-to-1 jet rate (middle) and the second photon transverse momentum
    (bottom).
    $\delta$ is varied logarithmically between $10^{-7}$ and $10^{-3}$.
    For 91.2\,GeV, the ratio is given with respect to $\delta=10^{-7}$
    while for 500\,GeV, $\delta=10^{-6}$ is used as the reference.
    \label{fig:results:varyDelta}
  }
\end{figure}

In Fig.\ \ref{fig:results:varyDelta}, we vary $\delta$ between $10^{-7}$
and $10^{-3}$ in isolation, keeping $\epsilon$ fixed at $10^{-8}$ as the
outcome of the previous validation exercise.
Since $\epsilon$ and $\delta$ enter the structure function rescaling
approximately through their ratio, this should be thought of as varying
the ratio between the two from $10$ to $10^5$.
We might expect that the behaviour could be pathological at both ends of
this spectrum: if the ratio is too small, the reweighting factor differs
significantly from unity, and if it is too large this implies a too-small
shower $x_\text{max}$ and a poor description of radiation.

On the top row of Fig.\ \ref{fig:results:varyDelta}, we show the
neutrino-pair invariant mass and verify that this observable is invariant
on the percent level under changing $\delta$.
The middle row, as before, shows the 0-to-1 jet rate.
For the 91.2\,GeV process, choices between $10^{-7}$ and $10^{-4}$ all
agree to within 1\%, but $\delta=10^{-3}$ does not adequately describe
the very soft/collinear radiation.
Similarly, the second photon $k_\mr{T}$ is well-described down to 10\,MeV
by the choice $\delta=10^{-4}$ and any smaller values.

However, at 500\,GeV, the variation of these radiative observables with
the choice of $\delta$ is larger.
In particular, we see that $\delta=10^{-7}$ no longer agrees with
$10^{-6}$, and in fact deviates from the established pattern for harder
radiation even than $\delta=10^{-3}$.
This is evidence that a ratio $\delta/\epsilon=10$ is insufficient for
good rescaling of the structure function at this energy, even using the
linear method.
For this reason, the ratio plots for $\sqrt{s}=500$\,GeV are shown with
respect to $\delta=10^{-6}$.

Focusing on the 0-to-1 jet rate, we find that $\delta=10^{-4}$ gives a
good description of the spectrum down to a transverse-momentum-equivalent
value of 30\,MeV.
The description of the second photon transverse momentum is of lower
quality for both collider energies, but the gain in shower speed is
significant enough to accept deviations on the order of 10\% in the
description of the second hardest photon (and only when this is
particularly soft).
A computational speedup allows the production of higher-quality results
overall for the same compute time, hence allowing the better description
of more relevant physical effects.
As a result, we choose $\delta=10^{-4}$ for the remainder of this paper.

\subsection{Validating the \texorpdfstring{\MCatNLO}{MC@NLO} implementation of the shower}

\begin{figure}
  \centering
  \includegraphics[width=0.45\textwidth]{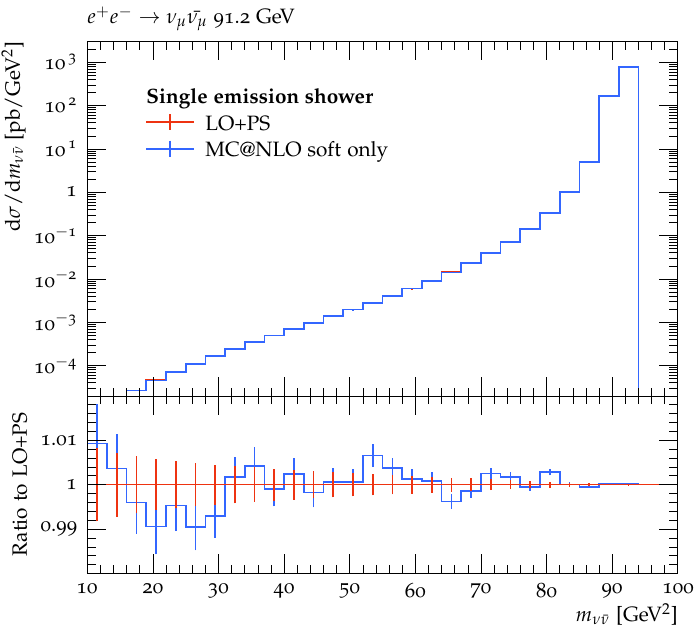}
  \includegraphics[width=0.45\textwidth]{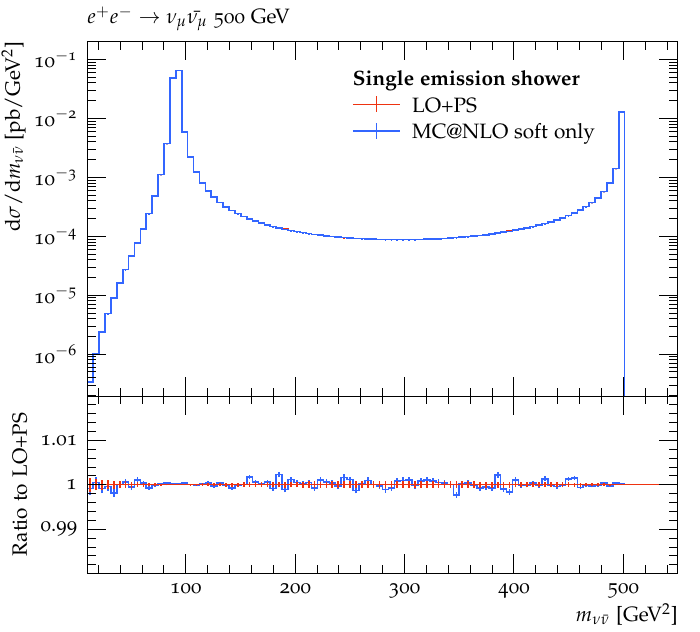}
  \hfill
  \includegraphics[width=0.45\textwidth]{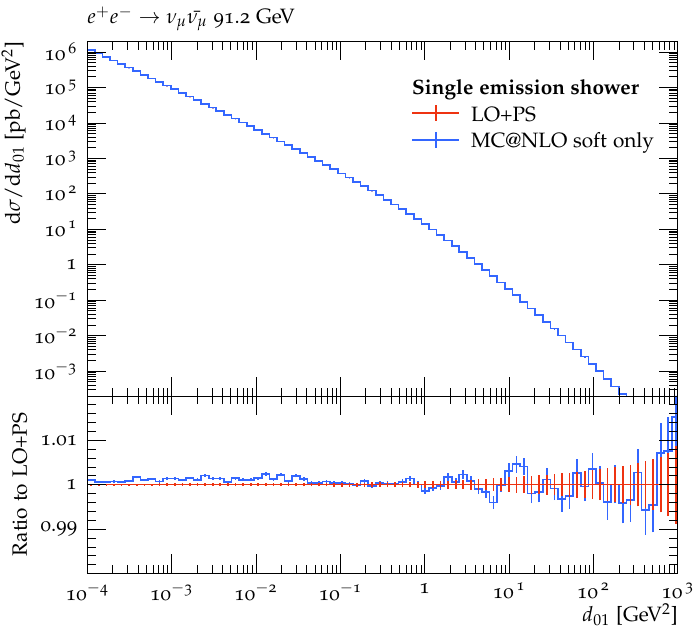} 
  \includegraphics[width=0.45\textwidth]{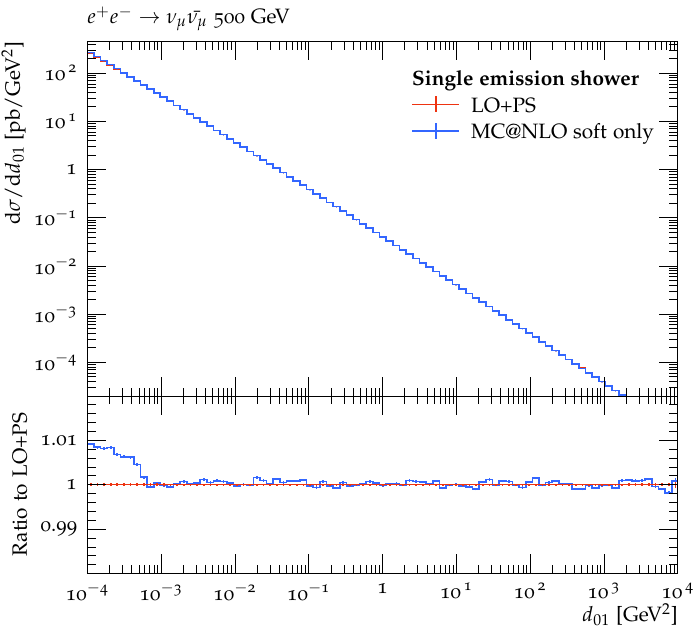}
  \caption{
    \MCatNLO without hard NLO corrections (see main text for details),
    compared with the LO parton shower prediction, for the process
    $e^+ e^- \to \nu_\mu \bar{\nu}_\mu$ at two different collider
    energies, 91.2\,GeV (left) and 500\,GeV (right).
    The observables shown are the neutrino-pair invariant mass (top)
    and the 0-to-1 jet rate (bottom).
    To illustrate the agreement, only one shower emission is allowed,
    since further emissions are described by the standard parton shower
    in both cases.
  \label{fig:results:validation}}
\end{figure}

To validate the implementation of the \MCatNLO method as compared to
the standard parton shower, we set $\tilde{\mr{V}}=-\sum \mr{I^S}$ and
neglect $\mr{D^A}-\mr{D^S}$, and expect that the $\mathbb{S}$-event
contribution is exactly equal to the LO standard shower contribution.
Fig.\ \ref{fig:results:validation} shows that they agree at the per-mille
level for the Born-level observable $m_{\nu \bar{\nu}}$, the neutrino-pair
invariant mass.
For the 0-to-1 jet rate, the agreement is 0.1\% when we study the process
at 91\,GeV, and at the same level for 500\,GeV for
$d_{01} > 10^{-3}\,\GeV^2$.
Noting that there is a 1\% difference between LO+PS and the
modified-$\mathbb{S}$ at the extremely soft end of the $d_{01}$ spectrum,
this study allows us to quantify the impact of cases where the overestimate
described in Sec.\ \ref{sec:methods:electronSF} fails to overestimate
the splitting function and the splitting then occurs.
When this happens, the \MCatNLO shower overweights the event, while the LO
shower does not.
The effect shown will have no meaningful physical effect and as
such we conclude that the overestimate is sufficient for our purpose.

\begin{figure}
  \centering 
  \includegraphics[width=0.4\textwidth]{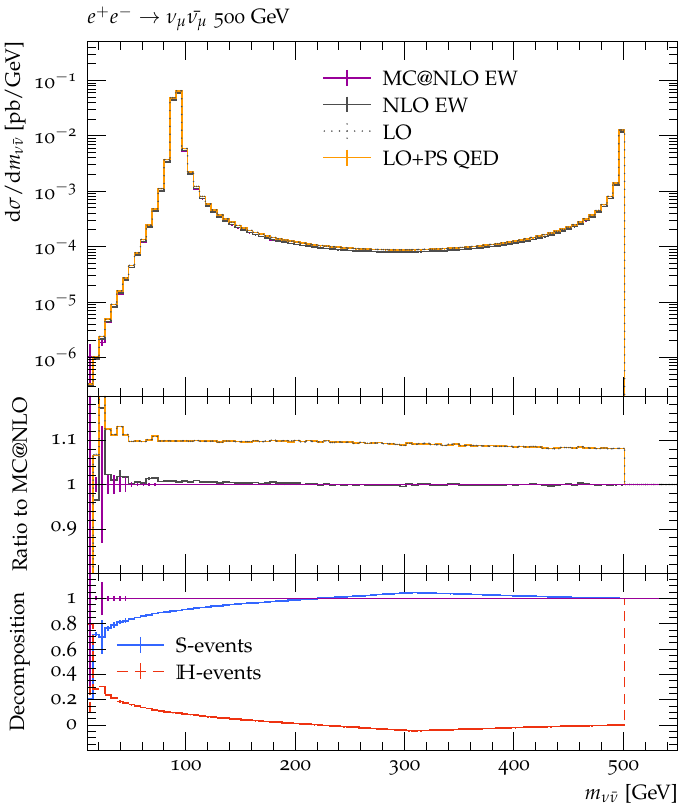}
  \includegraphics[width=0.4\textwidth]{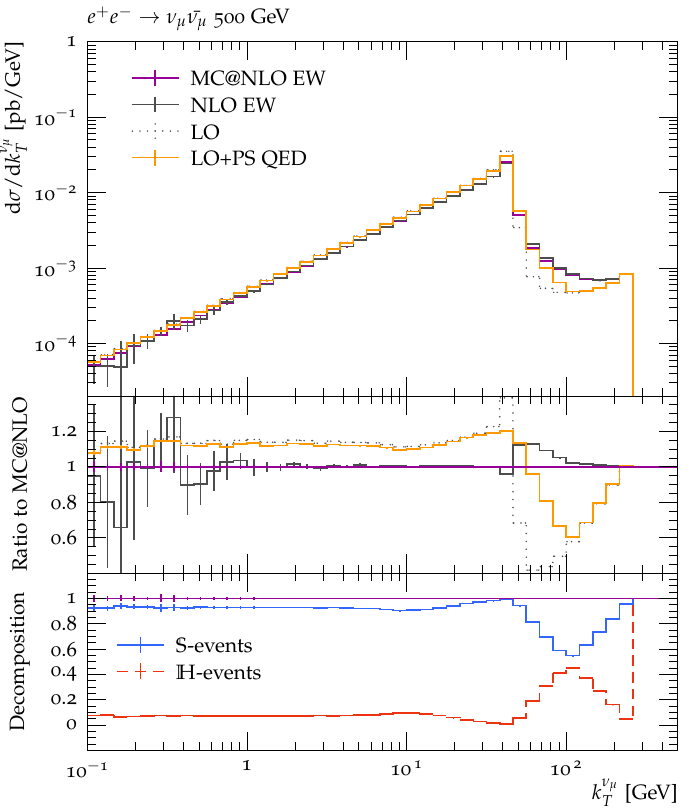}
  \hfill
  \includegraphics[width=0.4\textwidth]{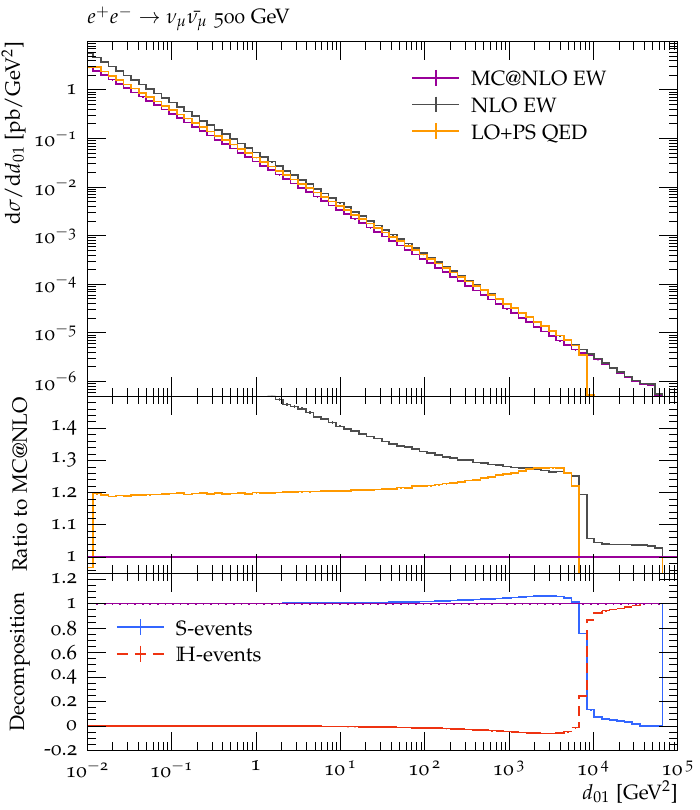}
  \includegraphics[width=0.4\textwidth]{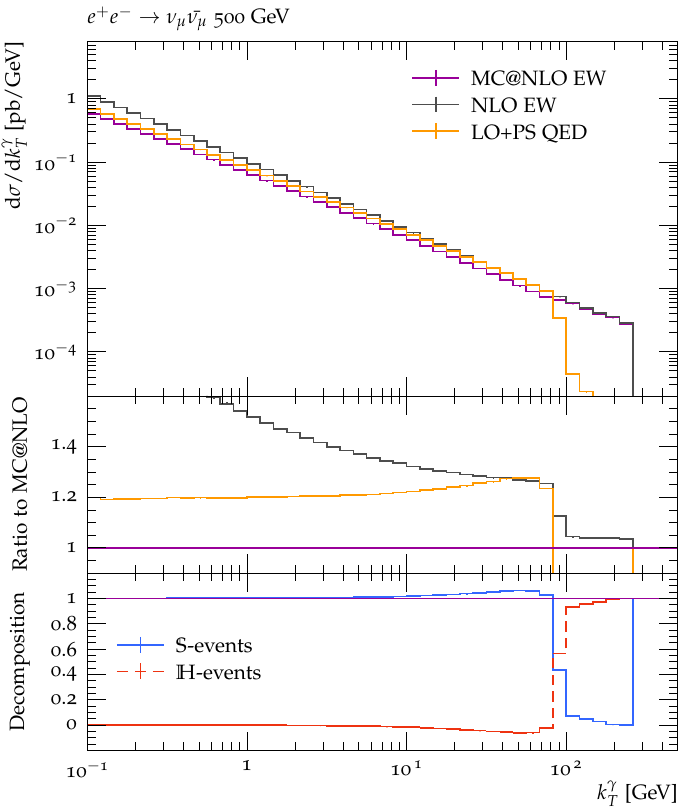}
  \hfill
  \includegraphics[width=0.4\textwidth]{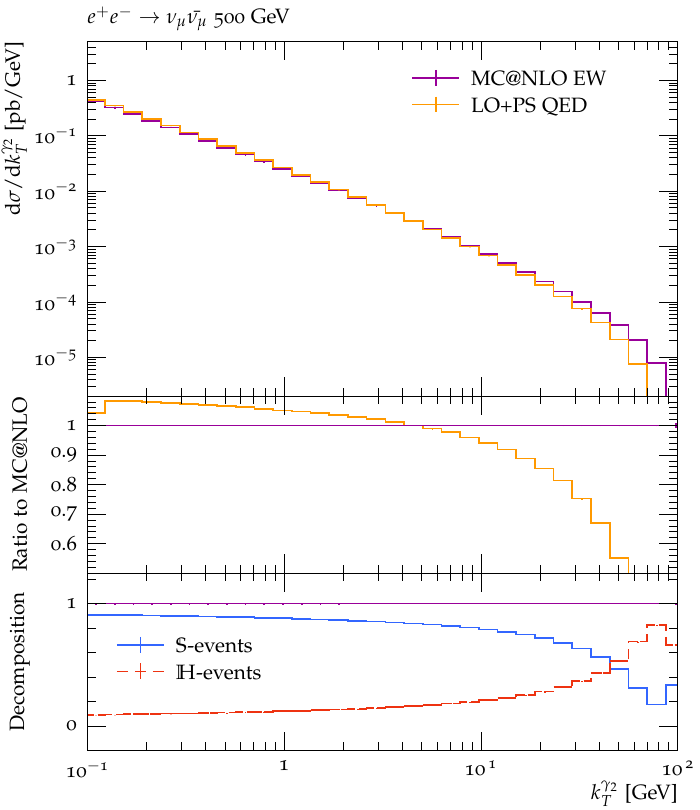}
  \includegraphics[width=0.4\textwidth]{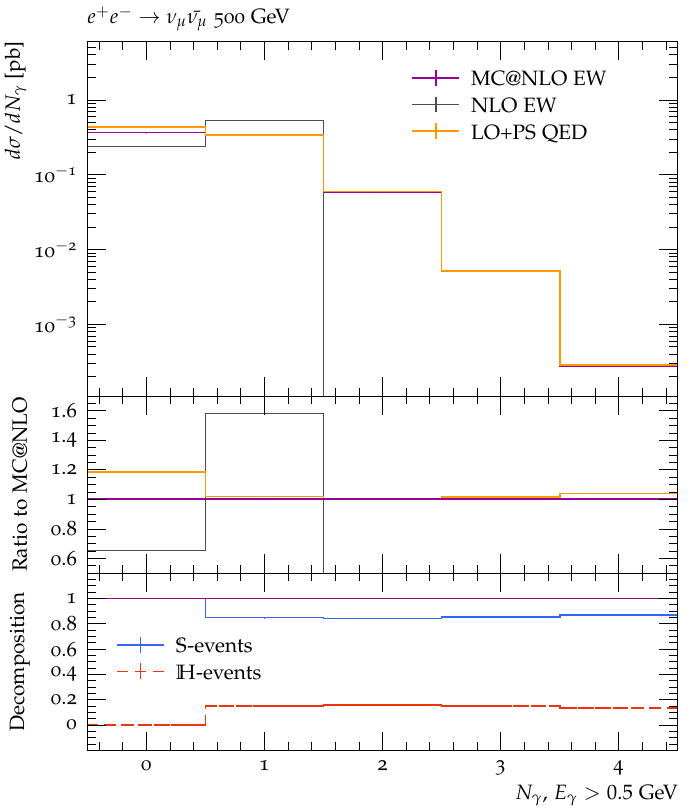}
  \caption{
    A comparison of \MCatNLO, the fixed-order NLO prediction and the
    leading-order prediction with and without the parton shower for
    the process $e^+ e^- \to \nu_\mu \bar{\nu}_\mu$ at 500\,GeV.
    The lower ratio plot in each subfigure shows the composition of
    the \MCatNLO prediction in terms of the $\mathbb{S}$- and
    $\mathbb{H}$-events.
    \label{fig:results:mcatnlo-general}
  }
\end{figure}

We also study the full \MCatNLO contribution and compare it to the LO
parton shower prediction and the fixed-order NLO prediction.
The total LO cross section is 0.837\,pb and the NLO cross section is
0.768\,pb: a $K$-factor of -8\%.
We find a scheme uncertainty, associated with the choice of mixed
$\alpha$ scheme described in Sec.\ \ref{sec:methods:ewscheme},
of 0.3\% on the total cross section for both fixed-order NLO and \MCatNLO,
based on the difference between the pure $G_\mu$ scheme cross section
and that obtained with the mixed scheme.
We find that the total cross section generated by the \MCatNLO method
agrees with the NLO one well within this scheme uncertainty, and show
the associated distributions in Fig.\ \ref{fig:results:mcatnlo-general}.

For clarity, as well as the total \MCatNLO result, we also show the
composition of the \MCatNLO prediction in terms of the
$\mathbb{S}$-events and $\mathbb{H}$-events in the lower ratio plot of
each subfigure.
We find that the $\mathbb{S}$-events are positive definite or zero for
all observables, whilst the $\mathbb{H}$-events contribute negatively
in some regions of phase space.

It is important to note that the initial-state parton shower does not
change the final-state invariant mass by construction, due to the
kinematic scheme \cite{Schumann:2007mg}.
The top-left subfigure shows the neutrino-pair invariant mass and its
ratio to the NLO and LO predictions.
From the upper ratio plot it is clear that the \MCatNLO and fixed-order
NLO agree below 1\% across the spectrum.
We also observe that the LO and LO+PS predictions lie on top of one
another, illustrating the same effect at LO.
We find that the effect of including NLO effects, on top of the LL
structure function, is a shift in the normalisation.

In contrast, in the top right plot of Fig.\ \ref{fig:results:mcatnlo-general}
we show the (matter) neutrino transverse momentum, which is sensitive
to the effects of a parton shower if hard emissions are kinematically
allowed.
We see that the LO and parton shower curves, and the NLO and \MCatNLO curves,
are each
consistent at small neutrino $k_\mr{T}$ but that the shower has a
sizeable effect on the neutrino transverse momentum above $m_Z/2$,
especially at LO where the effect is tens of percent.
On the other hand, compared to fixed-order NLO, the \MCatNLO correction
is smaller, of the order of 10\%.
In the case of \MCatNLO, we see that both the $\mathbb{S}$-events and
$\mathbb{H}$-events are positive and non-negligible across
the whole spectrum for this observable.

In the middle row of Fig.\ \ref{fig:results:mcatnlo-general}, we study
radiative observables: the 0-to-1 jet rate (left) and the transverse momentum
of the hardest photon.
Here we see the effect of the definition of $\mr{D^A}$ in our
version of the \MCatNLO method, see Eq.\ \eqref{eq:methods:smcatnlo},
and the scale choice of $\mu_Q^2 = m_{\nu \bar{\nu}}^2$ very clearly.
The hard-emission tail of both plots is dominated by the \MCatNLO
$\mathbb{H}$-events and the real-emission matrix element $\mr{R}$,
while the $\mathbb{S}$-events dominate at lower $k_\mr{T}$.
The transition between the $\mathbb{S}$-events and the $\mathbb{H}$-events
is gradual in general,
since it is determined event-by-event using the kinematic scale $\mu_Q^2$,
but in this case the presence of the $Z$ resonance
means that $\mu_Q$ is highly peaked around the $Z$ mass.
Therefore, above $m_Z$ there is an abrupt shift from a small and negative 
$\mathbb{H}$-event contribution to a
large positive one, since the parton shower kernels are switched off.
We see that the LO shower does not describe the spectrum in this region,
while the \MCatNLO curve agrees, within a few percent, with the fixed-order
NLO one.
We see that similarly to a QCD shower, the introduction of a
properly-matched parton shower to the NLO 
prediction leads to a levelling-off of the divergent soft end of the
radiation spectrum, but in QED, the position in $k_\mr{T}$ of the Sudakov peak
is much lower than for QCD \cite{Belloni:2026vnv} because
of weaker logarithms due to the smaller coupling.
A similar, but much smoother, scale-dependent transition between
$\mathbb{S}$- and $\mathbb{H}$-events is visible in the second-hardest
photon $k_\mr{T}$ distribution (bottom left).
This observable is described only at leading-logarithmic accuracy by both
the standard shower and \MCatNLO,
but differences are still seen due to the different kinematics of the
$\mathbb{S}$- and $\mathbb{H}$-events, both of which
undergo standard showering.

Finally, the photon multiplicity (bottom right) shows the cross section
differential in the number of photons with transverse momentum above a cut
$p_\mathrm{T}^\text{min}$, here 0.5\,GeV.
This observable differs only in a constant factor for $N_\gamma>0$ between
LO+PS, $\mathbb{S}$-events and the full \MCatNLO. The fixed-order NLO prediction has fewer unresolved-emission events than the
showered predictions.
The correction to the zero-photon bin from the $\mathbb{H}$-events is small,
since, by definition, below the 0.5\,GeV cut the parton shower describes the
real-emission matrix element very well.
In all resummed predictions, each extra photon occurs with a frequency of
around an order of magnitude less than the previous, consistent with a
factor proportional to $\alpha \log{(s/(p_\mathrm{T}^\text{min})^2)}$.

\section{\texorpdfstring{$ZH$}{ZH} production at a future \texorpdfstring{$e^+e^-$}{ee} collider} \label{sec:results}

The FCC-ee proposal sets forth the production of
$2.2\times 10^6$ $ZH$ events at $\sqrt{s}=240\,\GeV$
with $10.8 \,\text{ab}^{-1}$ of integrated luminosity \cite{FCC:2025lpp}.
Since the threshold for production of the on-shell $ZH$ state is
approximately 216\,GeV, there is somewhat restricted
phase space for hard quasi-collinear radiation in these events.
In this regime, the fixed-order correction is the most important
hard radiation correction, and the structure function takes care
of the soft and purely collinear radiation.
However, $3.7\times 10^5$ $ZH$ events are expected during the
$2.70\,\text{ab}^{-1}$ run above the $t\bar{t}$ threshold at 365\,GeV.
At this energy, there can be a large contribution from hard
quasi-collinear radiation as well as multiple semi-hard photons.
Therefore, we expect that the parton shower contribution, separately
from the NLO correction, will be important at this higher energy and
many observables will not be fully described by the structure function
alone.

We produce on-shell $ZH$ at NLO EW using the setup outlined in
Sec.\ \ref{sec:validation}, but set the widths of all unstable
particles to zero as they now occur as asymptotic states
in our calculation.
We continue to use the mixed $\alpha$ scheme outlined in section
\ref{sec:methods:ewscheme}.
For both the fixed-order and the showered predictions, we use the
electron structure function, Eq.\ \eqref{eq:methods:electronSF},
linearly rescaled near the singularity with $\epsilon = 10^{-8}$
and $\delta=10^{-4}$.
For the parton shower infrared cutoff we set $t_c=10^{-4}\,\GeV^2$
for $\sqrt{s}=240\,\GeV$ and $t_c=10^{-3}\,\GeV^2$ for
$\sqrt{s}=365\,\GeV$.
As before, we restrict the parton shower to allow only photon
emissions, to isolate their dominant effect on the relevant observables.
The decays of the $Z$ and Higgs can be included using the standard
\Sherpa decay module \cite{Hoche:2014kca}.
Nominally, QED corrections to these decays are accounted for
using the YFS soft-photon resummation \cite{Yennie:1961ad,Schonherr:2008av}
properly accounting for resonance shapes
\cite{Sherpa:2024mfk,Kallweit:2017khh}, while QCD
corrections are effected through the standard QCD parton shower
\cite{Schumann:2007mg}.
A completely off-shell treatment of intermediate resonances,
e.g.\ in a off-shell $e^+e^-\to \mu^+\mu^-\mr{b}\bar{\mr{b}}$
calculation, is easily possible at the fixed-order level at LO
and NLO EW, but the matching to a QED parton shower necessitates
a resonance-aware shower spectator assignment
to avoid distorting the resonance shapes \cite{Flower:2026lui}.

\subsection{Threshold production: \texorpdfstring{240\,GeV}{240 GeV}}

\begin{figure}
  \centering 
  \includegraphics[width=0.4\textwidth]{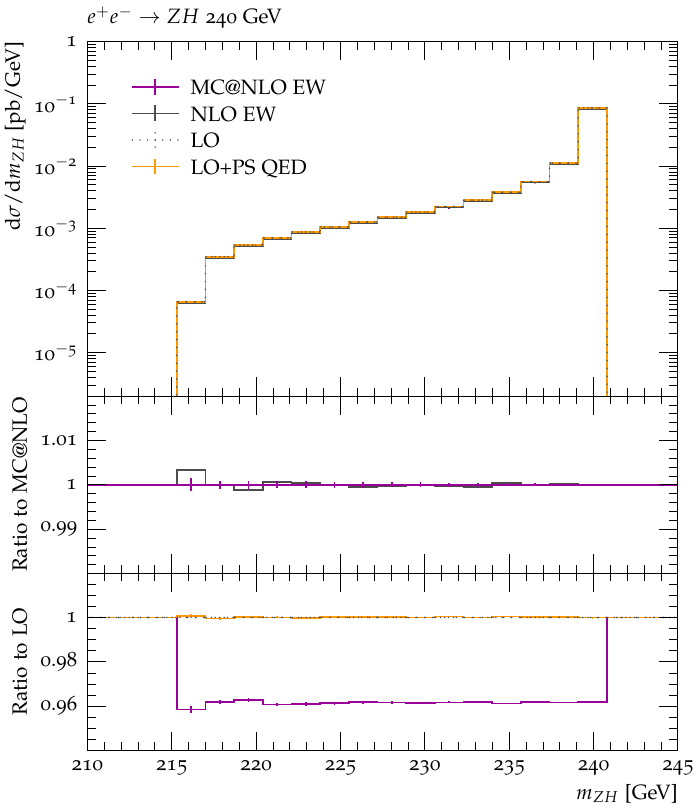}
  \includegraphics[width=0.4\textwidth]{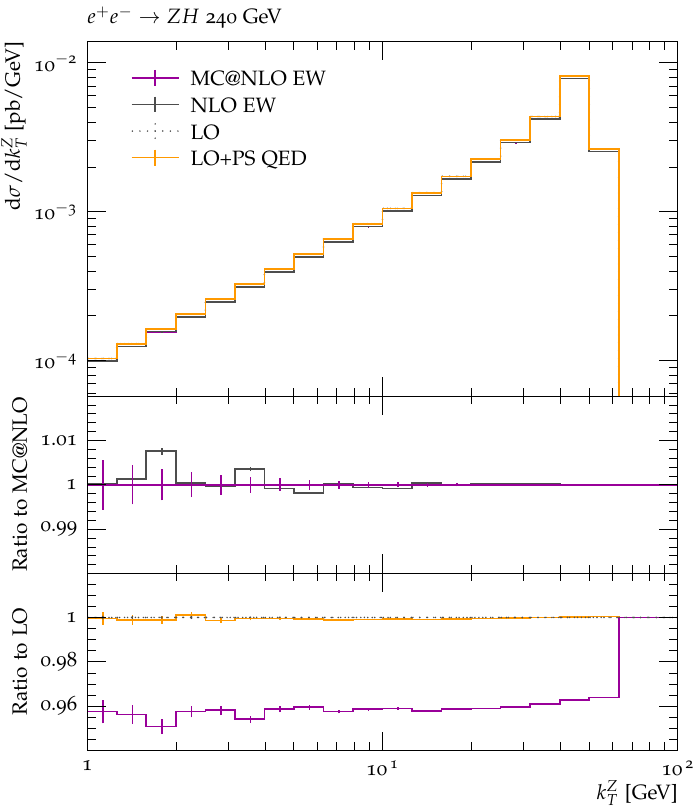}
  \hfill
  \includegraphics[width=0.4\textwidth]{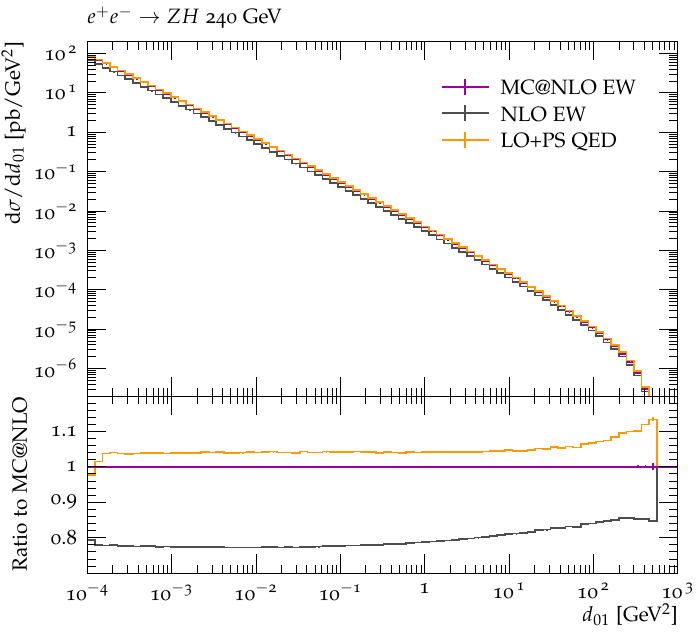}
  \includegraphics[width=0.4\textwidth]{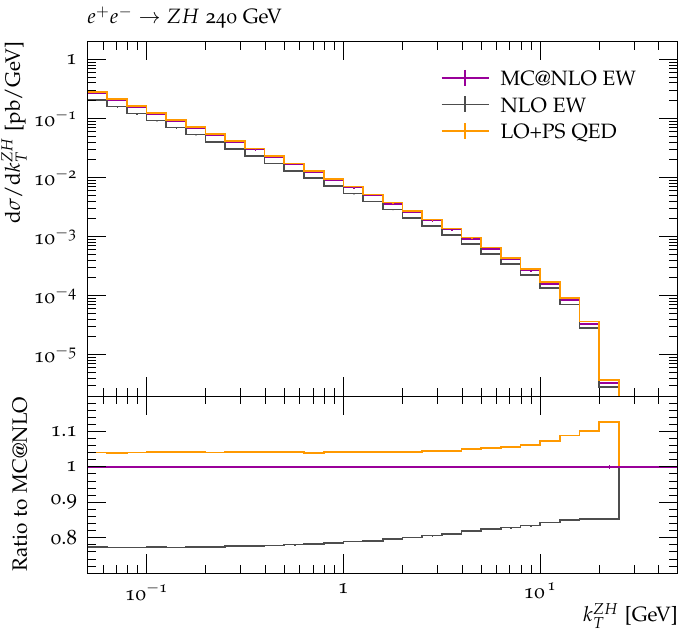}
  \hfill
  \includegraphics[width=0.4\textwidth]{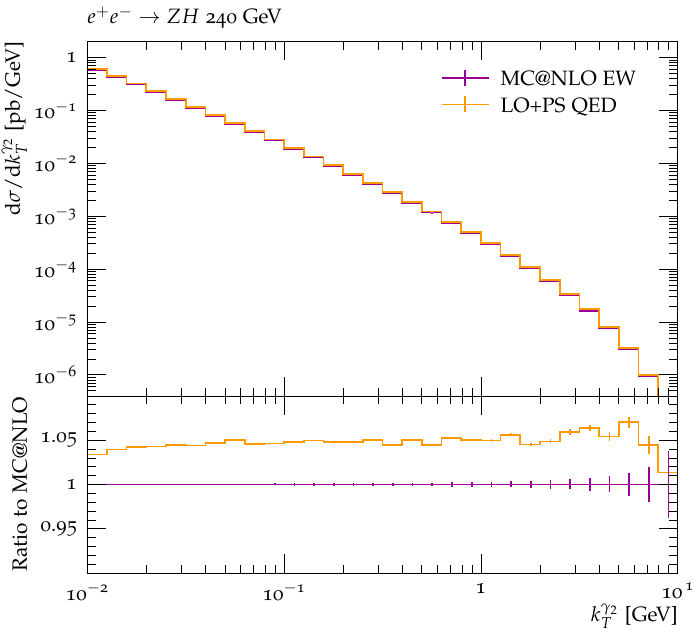}
  \includegraphics[width=0.4\textwidth]{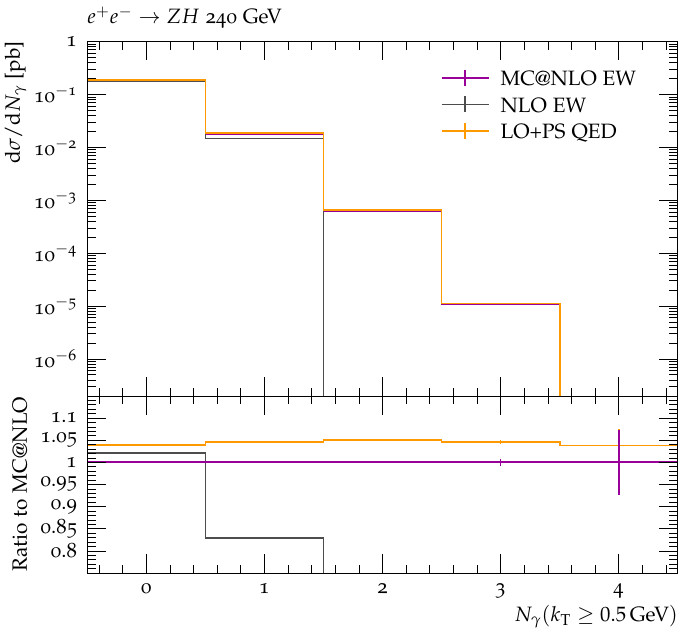}
  \caption{
    A comparison of \MCatNLO, the fixed-order NLO prediction and the
    leading-order prediction with and without the parton shower for
    the process $e^+ e^- \to ZH$ at 240\,GeV.
    \label{fig:results:mcatnlo-ZH240-general}
  }
\end{figure}

We begin our discussing of $ZH$ production by investigating
the properties of our implementation in threshold production.
In Fig.\ \ref{fig:results:mcatnlo-ZH240-general} we show the
EW \MCatNLO prediction for this process, the LO prediction
with and without the QED parton shower,
and the fixed-order NLO EW prediction.
The observables have been chosen reminiscent of those used in
the test case of neutrino production, but with a view to including
relevant experimental quantities.
For example, the $ZH$ final state is typically reconstructed
from its decay products, and this combined final state is then studied.
Here we have shown predictions for the invariant mass of the $ZH$
system and its overall transverse momentum
$k_\mathrm{T}^{ZH}$, in addition to the $Z$ transverse momentum.
It is clear from kinematic constraints that the latter is highly
correlated with the Higgs transverse momentum, and that the $ZH$
transverse momentum is highly
correlated with that of the hardest photon.

The $ZH$ invariant mass (top left subfigure of
Fig.\ \ref{fig:results:mcatnlo-ZH240-general}) shows, as before,
that the resummed predictions reproduce the fixed-order ones,
at their respective order, to sub-percent precision.
The correction to the total cross section is
-4\%, flat in the invariant mass distribution.
This result is exactly reproduced by our \MCatNLO computation.

In the top-right plot we show the $Z$ transverse momentum, plotted
logarithmically from 1\,GeV up to the kinematic limit.
Whilst the initial-state parton shower can in principle change
the distribution in this observable, we find that the effect is very
small due to the lack of phase space for hard collinear radiation,
both at LO and NLO.
This is in contrast to the case of neutrino production, where the
neutrino $k_\mr{T}$ received sizeable corrections from the inclusion
of parton shower effects.
The lower ratio plot shows that the NLO correction is almost flat
in this observable, again in contrast to high-energy neutrino
production.

We show radiative observables on the middle row of
Fig.\ \ref{fig:results:mcatnlo-ZH240-general}.
It can be seen that the transverse recoil of the $ZH$ final state
(middle right) is very highly correlated with the 0-to-1 jet rate
(middle left) under these conditions.
Again, it is important to note that like the invariant mass, these
observables, which vanish at Born level, will be changed significantly
by final-state radiation, and are thus described incompletely, serving
only to quantify the initial-state radiation which is our focus here.

It is clear that the parton shower approximation is an overestimate
of the photon-emission matrix element $\mr{R}$, as seen in the
comparison of the LO+PS jet rate compared with the NLO prediction.
The \MCatNLO prediction lies between the two.
Unlike for neutrino production in which the parton shower starting scale
was primarily identified with $t_n=m_{\nu\bar{\nu}}^2\approx m_Z^2$, there is
no dominant scale identification for the parton shower starting scale
$t_n=m_{ZH}^2$ in this process, hence there is no region in which the
$\mathbb{S}$-events do not contribute.
Truncated at relative $\mathcal{O}(\alpha)$, the NLO and \MCatNLO
predictions agree for large $k_\text{T}^{ZH}$ as shown in
Sec.\ \ref{sec:methods:mcatnlo}, but it is clear from the plot that
higher-order contributions are of the order of 15\% for these observables.
These originate from the smoothing behaviour of \MCatNLO,
in which virtual corrections are present within the radiative phase
space, but this behaviour is of $\order(\alpha^2)$ and can only
be constrained by a dedicated NNLO calculation.

We also show the properties of the resummation in \MCatNLO using
the second-hardest photon transverse momentum, $k_\mr{T}^{\gamma_2}$
in the lower left plot of Fig.\ \ref{fig:results:mcatnlo-ZH240-general}.
For this observable, the \MCatNLO correction compared to LO+PS is
roughly 5\% lower; however, the shapes are almost identical,
since the virtual contribution is flat.
Both the LO and \MCatNLO predictions are less than leading-order
in this observable---only the resummed terms contribute---but in
\MCatNLO these are resummed on top of the $\bar{\mr{B}}$ matrix element,
in contrast to LO+PS.

Finally, the photon multiplicity is shown in the bottom-right plot.
Unlike for neutrino production at 500\,GeV, we see fewer than 1 per
million events contain four or more non-soft photons.
The expected logarithmic decrease in $\done \sigma/\done N_\gamma$
in increasing $N_\gamma$ is amplified here by an additional drop due
to kinematic constraints.
Nevertheless, we find the expected relative 0- and 1-bin contributions at NLO and
in \MCatNLO. We also see that the ratio between the LO+PS and \MCatNLO predictions
flattens out from the 2-photon bin onwards, since both are determined by the same
standard QED parton shower.

\subsection{Radiative production: \texorpdfstring{365\,GeV}{365 GeV}}

\begin{figure}
  \centering 
  \includegraphics[width=0.4\textwidth]{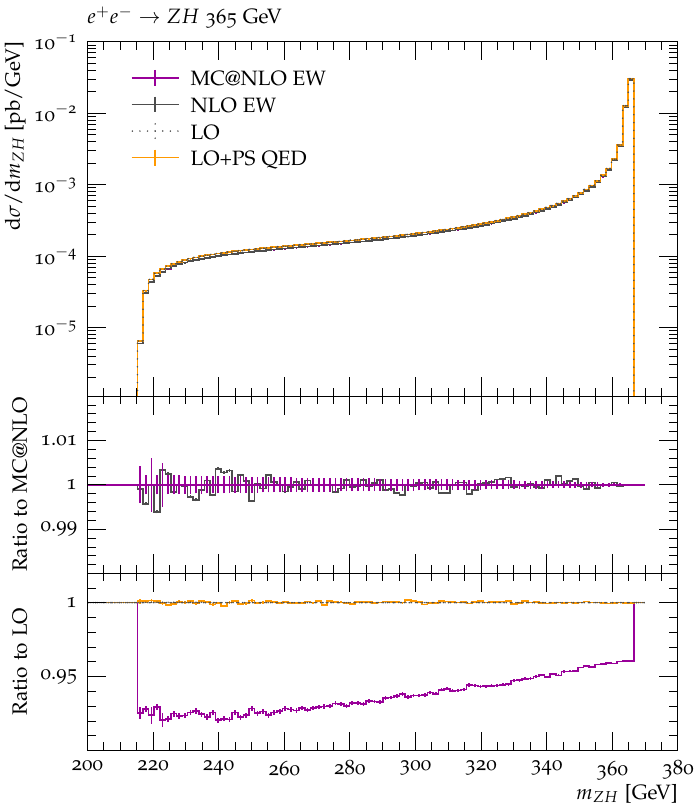}
  \includegraphics[width=0.4\textwidth]{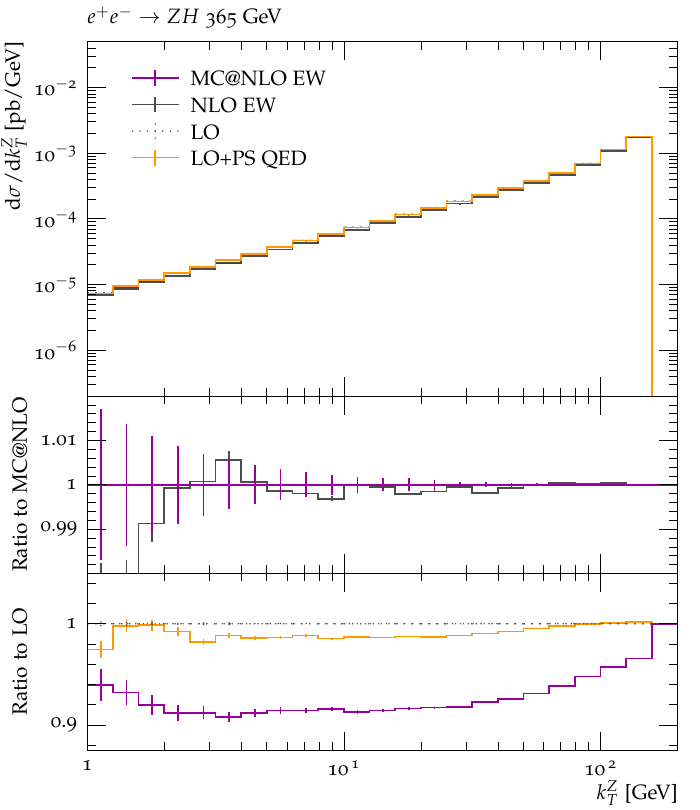}
  \hfill
  \includegraphics[width=0.4\textwidth]{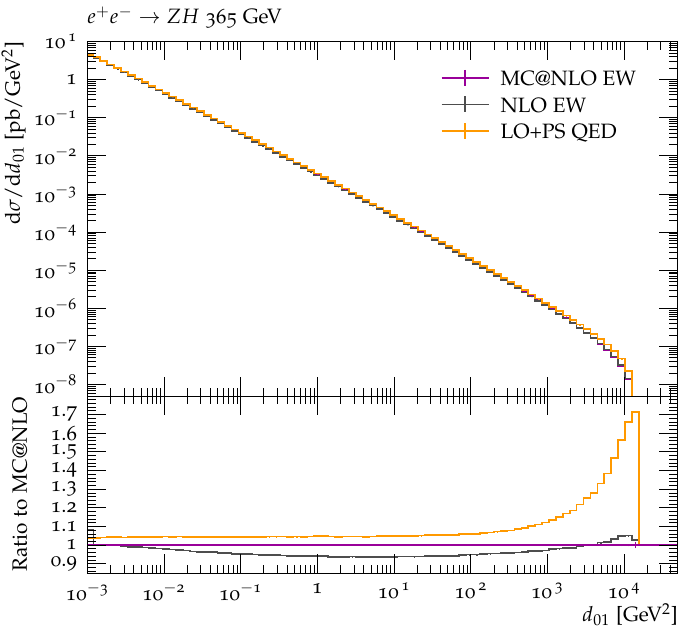}
  \includegraphics[width=0.4\textwidth]{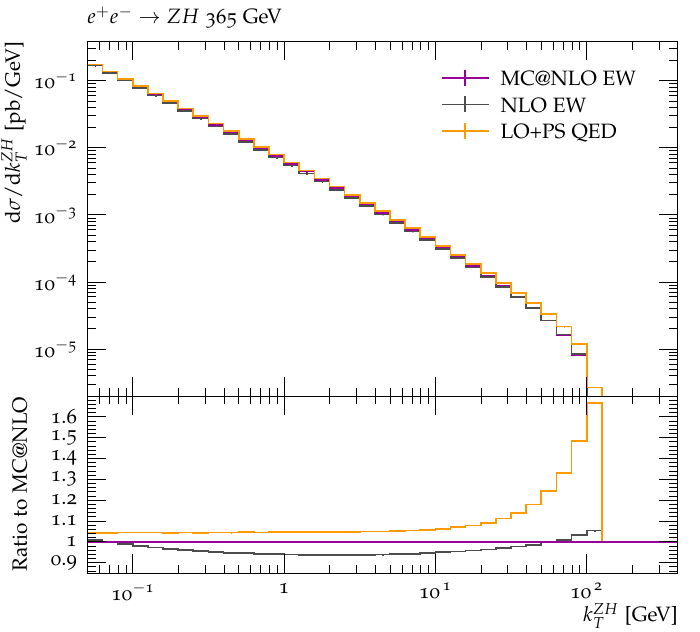}
  \hfill
  \includegraphics[width=0.4\textwidth]{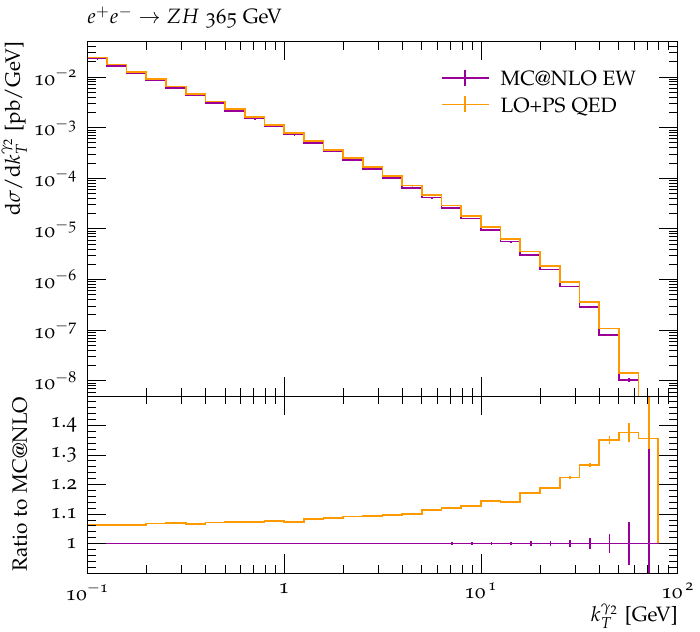}
  \includegraphics[width=0.4\textwidth]{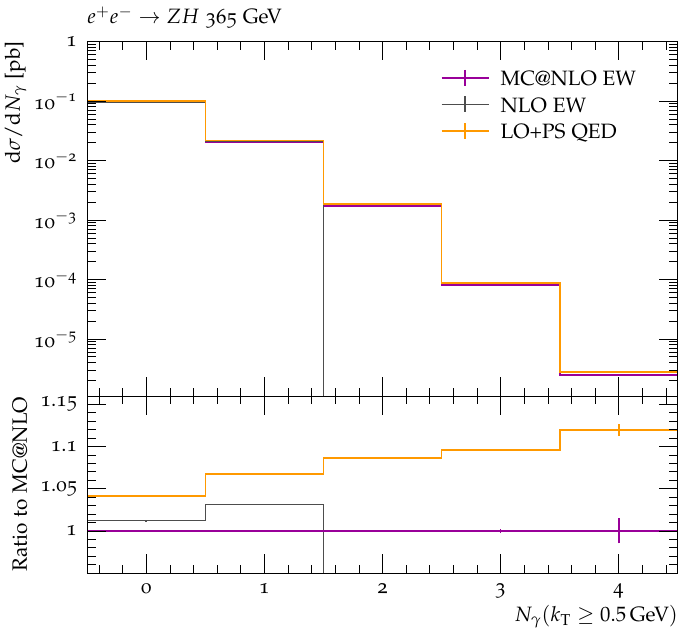}
  \caption{
    A comparison of \MCatNLO, the fixed-order NLO prediction and the
    leading-order prediction with and without the parton shower for
    the process $e^+ e^- \to ZH$ at 365\,GeV.
    \label{fig:results:mcatnlo-ZH365-general}
  }
\end{figure}

We now investigate the changes in the description of $ZH$ brought about
by the increased phase space available for QED radiation when the
centre-of-mass energy of the collision is increased.
To this end, in Fig.\ \ref{fig:results:mcatnlo-ZH365-general}, we show
the set of observables
defined above for the $\sqrt{s}=365\,\text{GeV}$ $e^+e^- \to ZH$ process.
As for 240\,GeV, the NLO and \MCatNLO predictions for the total cross
section and the $ZH$ invariant mass agree within 0.5\%, as expected.
The NLO EW correction to the total cross section is -4.4\%.

Studying the effect of the parton shower on the LO prediction, we see
from the top right plot of Fig.\ \ref{fig:results:mcatnlo-ZH365-general}
that the parton shower decreases the $Z$ transverse momentum distribution
by up to 2\%.
Whilst one might expect that a similar deviation would be seen at NLO,
the effect is instead much smaller, illustrating that the dominant
radiative effect on $k_\mr{T}^Z$ is captured at NLO and that effects beyond
NLO are small for this observable.

By looking at the radiative plots (middle row), we see that LO+PS
overestimates the hard radiation particularly severely at this energy,
an effect of over 70\% compared to \MCatNLO, whereas the effect at
240\,GeV was 12\%.
Since the \MCatNLO suffers less from this shape difference, and is
roughly flat with respect to the fixed-order NLO for hard radiation,
the effect of the two calculations yields an almost equivalent $k_\mr{T}^Z$
distribution.
As for 240\,GeV, the $\mathbb{H}$-event contribution in the radiative
observables is to correct the parton shower's region of poor
approximation in hard radiation.
Again, there is no region in which the $\mathbb{S}$-events do not
contribute, and as such the \MCatNLO and NLO predictions do not exactly
agree even for hard radiation due to effects beyond NLO.

The hard photon multiplicity, shown in the bottom-right plot
of Fig.\ \ref{fig:results:mcatnlo-ZH365-general}, shows qualitative
differences with the corresponding plot at 240 GeV. Here, the LO parton
shower prediction is not flat with respect to the \MCatNLO one, but
instead predicts successively more photons. For both resummed predictions,
the drop-off in the number of photons is not as severe as for 240 GeV.
Interestingly, this results in the fixed-order NLO prediction lying
entirely above the \MCatNLO one in the 0-photon and 1-photon bins.
In all predictions, as before, an event is most likely to have zero
hard photons due to the large production threshold of this process.

It is interesting to note that despite the $\sqrt{s}=240\,\text{GeV}$
and 365\,GeV setups allowing for different radiation patterns in
principle, the shapes of all NLO predictions for the observables
studied here are very similar between the two different
centre-of-mass energies.
The higher centre-of-mass energy ultimately results in an extension
of the invariant mass peak seen in Fig.\ \ref{fig:results:mcatnlo-ZH240-general}
(top left) to a sharper peak at the higher energy.
This increases the size of the parton shower correction at LO, but
decreases the size of the correction at NLO.
Clearly, the dominant hard collinear radiation is already present
in the fixed-order NLO prediction (both from the matrix element and
from the structure function).
This suggests that if exclusive description of multiple photons and
generation of unweighted events is not required, it is sufficient to
consider NLO with a structure function at this accuracy.
Nonetheless, the features of \MCatNLO are desirable in a wide range
of processes.

The presence of phase-space cuts will also affect this assessment.
While a structure function can appropriately shift the momenta of
the initial-state leptons to capture the radiative return phenomenon,
only a QED parton shower (matched to NLO) adequately describes the
transverse boost which a final state may acquire due to initial-state
non-collinear radiation.
In the analyses presented here, we have been inclusive of such boosts,
but in general the effect on observables will be non-trivial due to
selection cuts.

\section{Conclusions}
\label{sec:conclusions}

In this paper we have presented the first automated,
process-independent matching of
a QED parton shower and an NLO EW calculation for lepton colliders.
In particular, the results shown here constitute the first
\MCatNLO EW prediction for on-shell $ZH$ production at proposed working points
at a future electron-positron collider, including the full NLO EW
corrections as well as the leading-logarithmic corrections from
the QED parton shower.
The method is general and, in its current form, is easily applicable
to all processes for which initial-final interference beyond NLO is
suppressed and which do not contain a combination of tagged (resolved)
and untagged (unresolved) photons at leading order.
Even considering these caveats, this includes a majority of colourless
processes at current and future electron-positron colliders.
To describe coloured final states, or for processes at hadron colliders,
the combined QCD+EW matched interleaved parton shower proceeds
following the methods presented here and in earlier \MCatNLO literature.
Preliminary results for the interleaved shower have been presented for
$gg\to H\to e^+e^-\mu^+\mu^-$ \cite{Flower:2024cpj}, and a detailed study
of the NLO QCD+EW matching to the QCD+QED parton shower has been presented
in \cite{Flower:2026lui}.
At low energies where perturbative QCD is not applicable, various
phenomenological models for describing hadronic cross sections can
be combined with the QED parton shower, and the matching proceeds
similarly.
In particular, the methods presented here are highly applicable
to the description of low-energy electron-positron collisions,
both for the energy scan and radiative return methods.
Bhabha scattering requires an excellent description of initial-final
interference, but di-muon and di-pion production (and their radiative
analogues) can be considered either in the presence or absence of
this interference.
The study of these processes will be the subject of future work.

The QED parton shower and the associated \MCatNLO EW have each
undergone stringent validation testing, detailed in section
\ref{sec:validation}.
To validate the methods, we have varied the parameters which control
infrared limits of different parts of the calculation, verified that
they do not change Born-level observables and that their impact on
radiative observables is predictable and small.
Based on these studies, we have chosen an optimal set of parameters
for a future collider setup.
In general, we expect that the optimal parameter choices will depend
on the order of magnitude of the centre-of-mass energy,
as well as the observables under consideration and fiducial cuts.

The main results which we have used to illustrate the predictions of
the method are for the $ZH$ production process at a future $e^+ e^-$
collider such as the FCC-ee.
We have shown how the \MCatNLO method reproduces the fixed-order
accuracy for the total cross section, the final-state invariant mass
and the $Z$ boson transverse momentum; and that it models soft and
collinear radiation beyond NLO according to the parton shower approximation.
In particular, there is a smooth transition between these regions
owing to the construction of the method and the choice of scales.

A key part of the method presented here is the automated identification
of $n$-parton configurations, and the scales associated with these, in
order to correctly identify a hardest emission in each event and to set
the shower starting scale accordingly.
This ensures the respective accuracy of the NLO correction and the
resummation.
We have achieved this identification using a QED-flavour-aware version
of \Sherpa's \MEPS scale setter.
In complement to this, when analysing the results, we used a
flavour-aware jet clustering algorithm to quantify the behaviour of the
parton shower.

In addition to the work directly on the \MCatNLO and parton shower,
it was necessary to modify the way that the electron structure function
is conventionally modelled in event generators.
Due to the parton shower's rescaling of initial-state momenta, the
traditional choice of rescaling the structure function near the
$x\to 1$ integrable singularity was not suitable for the LO parton shower,
let alone for the \MCatNLO method.
In this paper we have introduced a linear rescaling, which keeps the
structure function continuous until its numerical cutoff, and a
gradiated shower cutoff $x_\text{max}$.
This means that the parton shower can emit into the rescaled region,
but is then forbidden from acting again on that initial-state particle.
Ideally, however, the rescaling used would render the structure function
differentiable in the domain $[0,1-\epsilon)$.
For this purpose, we are considering sigmoid weighting functions.

Throughout this paper, we have used an electron structure function
which is derived starting from LO evolution kernels (the
Altarelli-Parisi functions) using LO initial conditions, $f_e(x,m_e^2)=\delta(1-x)$.
To go beyond this approximation, one can produce a next-to-leading
logarithmic structure function
\cite{Frixione:2019lga,Bertone:2019hks,Frixione:2021wtz,Bertone:2022ktl,
  Stahlhofen:2025hqd,Schnubel:2025ejl}.
This entails using NLO initial conditions, which preclude neglecting
the photonic initial state as we have done here.
A future plan for the QED parton shower in \Sherpa is to allow for
the evolution starting from a NLL structure function, or indeed from
a full parton distribution function, in analogy with QCD.
Lepton PDFs suitable for use in an event generator are still under
development, but are expected in the near future.

All analyses and plots were made using \Rivet \cite{Buckley:2010ar,Bierlich:2019rhm,Bierlich:2024vqo}.

\subsection*{Acknowledgements}

The authors would like to thank Alan Price, Carlo Carloni Calame,
and Stefan Dittmaier for helpful discussions.
Joanne Roper's careful reading of the manuscript is hugely appreciated.
MS is funded by the Royal Society through a University Research Fellowship
(URF\textbackslash{}R\textbackslash{}231031 and
 URF\textbackslash{}R1\textbackslash{}180549) and Enhancement Awards
(RF\textbackslash{}ERE\textbackslash{}210397,
 RGF\textbackslash{}EA\textbackslash{}181033 and
 CEC19\textbackslash{}100349).
LF is supported by the Leverhulme Trust, LIP-2021-014.
LF also acknowledges support from an STFC studentship under the STFC training grant 
ST/P001246/1 for the early parts of this work.

\appendix
\section{KP terms in the presence of lepton structure functions}
\label{app:kp}

This appendix details modifications to the Catani-Seymour
subtraction of QED singularities \cite{Catani:1996vz,Catani:2002hc,
  Schonherr:2017qcj} necessary to be used in conjunction with
lepton PDFs, namely the $\Kop$ and $\Pop$ operators in particular.
As a first observation, while the $\Kop$ operator contains a
factorisation-scheme dependent contribution, any scheme
dependence vanishes to LL accuracy \cite{Frixione:2021wtz}.
In practice, however, the \MSbar factorisation scheme leads
to large new contributions as $z\to 1$, and we, hence, adopt
the $\Delta$ scheme of \cite{Frixione:2021wtz} preserving
the LL structure to NLO.
This leaves us to deal with the numerical instabilities induced by the
integrable singularity at $z=1$ in the electron structure function,
see Sec.\ \ref{sec:methods:electronSF}.
Following the notation of \cite{Schonherr:2017qcj},
the $\Kop$ and $\Pop$ operator contributions to an arbitrary
IR-safe observable $O$ for beam $A$ can be expressed in the
following form
\begin{equation}
  \begin{aligned}[b]
    \langle O\rangle_{\KPop}^{A}
    =
    \sum_{a,b}\int\!\done\eta_a\done\eta_b\!
      \int\!\done\Phi_m
        \sum_{a'}\!\int\limits_0^1\!\done x_a\;
        \mr{B}_{a'b}(x_a\eta_a,\eta_b;\Phi_m)
	\bigg[
	  \Kop_{aa'}(x_a;\{\alphadip\})
	  \!+\!\Pop_{aa'}(x_a;\mu_F^2)
	  \vphantom{\Phi_m}
	\bigg]
    \,  O(\Phi_m)\!\!
  \end{aligned}
\end{equation}
where $a$ and $b$ would run over all possible partonic initial
states, although when using the usual electron structure functions,
$a,b\in\{e^+\}$ or $\{e^-\}$ only, depending on the beam.
$\eta_a$ and $\eta_b$ are their momentum fractions.
Likewise, $a'$ would run over all possible partonic initial
states, but is restricted to $\{e^+\}$ or $\{e^-\}$, respectively.
The hard scattering process is defined as $\mr{B}_{ab}=f_{a} f_b\;\mc{B}_{ab}$ to include the parton densities $f_a$ and $f_b$, and the partonic
cross section $\mc{B}_{ab}$.
$x_a\eta_a$ is then the remaining momentum fraction of flavour $a'$ after the integrated-over emissions $a\to a'+X$, which enters the hard scattering.
The $\Kop$ and $\Pop$ operators, resulting from
the combination of the integrated subtraction terms and the collinear
counterterms, depend on the additional $x$ integration variable as
\cite{Gleisberg:2007md,Schonherr:2017qcj}
\begin{equation}
  \begin{split}
    \left[g(x)\right]_++\delta(1-x)\,h(x)+k(x)\;.\nnb
  \end{split}
\end{equation}
Thus, using the definitions of the plus prescription and $\delta$-function,
{\allowdisplaybreaks
\begin{equation}\label{eq:app:kp:KPdef}
  \begin{split}
    \lefteqn{\hspace*{-5mm}\langle O\rangle_{\KPop}^{A}\vphantom{\int_A^B}
    }\\[-3mm]
    =\;&
      \sum_{a,b}\int\done\eta_a\done\eta_b
      \int\done\Phi_m\;
      \mr{B}_{ab}(\eta_a,\eta_b;\Phi_m)\\
    &{}
      \times\sum_{a'}
      \left\{\;
	\int\limits_{\eta_a}^1\done x_a
	\left[
	  \frac{f_{a'}\!\!\left(\tfrac{\eta_a}{x_a}\right)}{x_a f_a(\eta_a)}
	  \left(g^{aa'}(x_a)+k^{aa'}(x_a)\right)
	  -\frac{f_{a'}(\eta_a)}{f_a(\eta_a)}\,g^{aa'}(x_a)
	\right]
      \right.\\
    &\hspace*{12mm}
      \left.{}\vphantom{\frac{f_{a'}\left(\frac{\eta_a}{x_a}\right)}{x_af_a(\eta_a)}}
	+\frac{f_{a'}(\eta_a)}{f_a(\eta_a)}
	 \left(
	   h^{aa'}-G^{aa'}(\eta_a)
	 \right)
      \right\}
      \;O(\Phi_m)\;,
  \end{split}
\end{equation}
where, after factorising out the $x_a=1$ piece
of the hard scattering process, we have split the $x_a$ integration
into two regions: $x_a\in (0,\eta_a)$ and $x_a \in (\eta_a,1)$.
Integrating the remaining piece using $f_a(z)=0$
for $z>1$ and defining
$G^{aa'}(\eta)=\int_0^\eta\done x\;g^{aa'}(x)$ and
$h^{aa'}\equiv h^{aa'}(1)$.
When using electron structure functions, i.e.\ $a,a'\in\{e^+\}$ and
$b\in\{e^-\}$ only (and the reverse which follows trivially),
and wanting to execute the remaining integration over $x_a$
numerically, this form still suffers from the integrable
singularity of $f_e(z)$ as $z\to 1$.
We thus again introduce an $\varepsilon$-environment,
with $\varepsilon\ll 1$, to split the integration region
into two domains, $[\eta_{a},\eta_{a}/(1-\varepsilon)]$
and $[\eta_{a}/(1-\varepsilon),1]$,
\begin{equation}\label{eq:app:kp:KPeps}
  \begin{split}
    \lefteqn{\hspace*{-5mm}\langle O\rangle_{\KPop}^A
    \equiv
    \left.\langle O\rangle_{\KPop}^A\right|_{(\varepsilon=0)}}\\
    =\;&
      \sum_{a,b}\int\done\eta_a\done\eta_b
      \int\done\Phi_m\;
      \mr{B}_{ab}(\eta_a,\eta_b;\Phi_m)\\
    &{}
      \times\sum_{a'}
      \left\{\;
	\int\limits_{\frac{\eta_a}{1-\varepsilon}}^1\done x_a
	\left[
	  \frac{f_{a'}\!\!\left(\tfrac{\eta_a}{x_a}\right)}{x_a f_a(\eta_a)}
	  \left(g^{aa'}(x_a)+k^{aa'}(x_a)\right)
	  -\frac{f_{a'}(\eta_a)}{f_a(\eta_a)}\,g^{aa'}(x_a)
	\right]
      \right.\\
    &\hspace*{12mm}
      \left.{}\vphantom{\frac{f_{a'}\left(\frac{\eta_a}{x_a}\right)}{x_af_a(\eta_a)}}
	+\frac{f_{a'}(\eta_a)}{f_a(\eta_a)}
	 \left(
	   h^{aa'}-G^{aa'}(\eta_a)
	 \right)
      \right\}_{\varepsilon=0}
      \;O(\Phi_m)\\
    =\;&
      \left.\langle O\rangle_{\KPop}^A\right|_{\varepsilon}
      +
      \sum_{a,b}\int\done\eta_a\done\eta_b
      \int\done\Phi_m\;
      \mr{B}_{ab}(\eta_a,\eta_b;\Phi_m)
      \\
    &{}\hspace*{20mm}
      \times\sum_{a'}
      \left\{\;
	  \frac{g^{aa'}(\eta_a)+k^{aa'}(\eta_a)}{f_a(\eta_a)}
    \int\limits_{\eta_a}^{\frac{\eta_a}{1-\varepsilon}}
    \done x_a
	\left[
	  \frac{1}{x_a}\,f_{a'}\!\!\left(\tfrac{\eta_a}{x_a}\right)
	\right]
	  -\frac{f_{a'}(\eta_a)}{f_a(\eta_a)}\,\frac{\varepsilon\eta_a}{1-\varepsilon}\,g^{aa'}(\eta_a)
      \right\}
      \;O(\Phi_m)\;.
      \hspace*{-20mm}
  \end{split}
\end{equation}
While the $x_a$ integration can still be performed
numerically on the $[\eta_{a}/(1-\varepsilon),1]$
domain, giving $\left.\langle O\rangle_{\KPop}^A\right|_{\varepsilon}$,
both $g^{aa'}(x_a)$ and $k^{aa'}(x_a)$ are sufficiently
well behaved so we can integrate the remaining
piece on the small $\varepsilon$-environment analytically.
For $a=a'=e$, using the asymptotic form of the electron
structure function of Eq.\ \eqref{eq:methods:electronSF_limit} and
introducing a change of variables $\tfrac{\eta}{x}=1-\xi$,
the integral over the structure function simplifies to
\begin{equation}
  \int\limits_\eta^{\frac{\eta}{1-\varepsilon}}
  \frac{\done x}{x}\,f_e\left(\tfrac{\eta}{x}\right)
  \,\stackrel{\frac{\eta}{x}\to 1}
             {\relbar\joinrel\relbar\joinrel\longrightarrow}
    \beta
    \int\limits_0^\varepsilon
    \frac{\done \xi\,\xi^{\beta-1}}{1-\xi}
  \,\stackrel{\varepsilon\to 0}
             {\relbar\joinrel\relbar\joinrel\longrightarrow}
    \varepsilon^{\beta}\;.
\end{equation}
This gives us the final expression for the $\KPop$-terms for
the use of the electron structure function at eletron-positron
colliders,
\begin{equation}
  \begin{aligned}[b]
    \lefteqn{\hspace*{-5mm}\langle O\rangle_{\KPop}^{e^+}}\\
    =\;&
      \left.\langle O\rangle_{\KPop}^{e^+}\right|_{(\varepsilon)}
      +
      \int\done\eta_{e^+}\done\eta_{e^-}
      \int\done\Phi_m\;
      \mr{B}_{e^+e^-}(\eta_{e^+},\eta_{e^-};\Phi_m)\\
    &{}\hspace*{20mm}
      \times
      \left\{\;
        \frac{\varepsilon^{\beta}}{f_{e^+}(\eta_{e^+})}
        \left(g^{e^+e^+}(\eta_{e^+})+k^{e^+e^+}(\eta_{e^+})\right)
        -\varepsilon\eta_{e^+}\,g^{e^+e^+}(\eta_{e^+})
	  \right\}
      \;O(\Phi_m)
	  \;.
  \end{aligned}
\end{equation}
Particular care has to be taken if $\eta_a>1-\varepsilon$.
In this situation, a redefinition of
$\varepsilon\to\tilde{\varepsilon}=1-\eta_a$ is necessary.

It remains to choose a value for the regulator $\varepsilon$.
While $\varepsilon$ must not be to large for our approximation
in the analytical integration to hold, a large $\varepsilon$
also reduces the number of phase space points for which the
numerical integral is evaluated, increasing its Monte-Carlo
statistical uncertainty.
Likewise, for values of $\varepsilon<\delta$, as it is defined
in Sec.\ \ref{sec:methods:electronSF}, the reweighting from
$f_e\to W_e$ must not be applied in the numerator PDF ratios
from Eq.\ \eqref{eq:app:kp:KPdef} but still in their denominators.
However, for \MCatNLO, the value of
$\varepsilon$ impacts the phase-space generation in a broader way, since all Born-type events are 
generated together. A too-small $\varepsilon$ is therefore prohibitively inefficient. 
For the optimal efficiency, and minimal error, for both fixed-order and \MCatNLO we thus 
choose $\varepsilon=10^{-3}$.

\section{Splitting functions}
\label{app:sfs}


\begin{table}[t!]
  \centering
  \begin{tabular}{|c|c|c||c|c|c|}
    \hline
    \hl & $y$ & $z$ && $y$ & $z$ \\\hline
    FF\Hl
    & $\dst\hphantom{-}\frac{p_ip_j}{p_ip_j+p_ip_k+p_jp_k}$ 
    & $\dst\frac{p_ip_k}{p_ip_k+p_jp_k}$
    & IF
    & $\dst\frac{p_ap_j}{p_ap_j+p_ap_k}$ 
    & $\dst\frac{p_ap_j+p_ap_k-p_jp_k}{p_ap_j+p_ap_k}$ \\\hline 
    FI\Hl
    & $\dst\frac{p_ip_j-\frac{1}{2}\left(m_\ijt^2-m_i^2-m_j^2\right)}{p_jp_a+p_ip_a}$ 
    & $\dst\frac{p_ip_a}{p_ip_a+p_jp_a}$
    & II
    & $\dst\frac{p_jp_a}{p_ap_b}$ 
    & $\dst\frac{p_ap_b-p_jp_a-p_jp_b}{p_ap_b}$ \\\hline 
  \end{tabular}
  \caption{
    Definitions of the dipole variables $y$ and $z$
    for all four dipole configurations.
    \label{table:methods:yz}
  }
\end{table}

In this appendix we detail the splitting functions used in the QED
parton shower and \MCatNLO.
To describe the QED splitting functions, we will need to introduce the
$\mr{U}(1)$ analogue to the colour correlator, the
charge correlator
\cite{Yennie:1961ad,Dittmaier:1999mb,Dittmaier:2008md,
  Kallweit:2017khh,Schonherr:2017qcj}, defined as
\begin{equation} \label{eq:app:charge_correlator}
  \begin{split}
    \mf{Q}_{\ijt\kt}^2 =
    \begin{cases}
      \frac{Q_\ijt \theta_\ijt Q_\kt \theta_\kt}{{Q_\ijt}^2} &
      \text{for } \ijt \neq \gamma \\
      \kappa_{\ijt\kt} &
      \text{for } \ijt = \gamma
    \end{cases}
    \qquad\qquad\text{with}\qquad\qquad
    \sum_{\kt \neq \ijt} \kappa_{\ijt\kt} = -1 \quad \forall \ijt=\gamma,
  \end{split}
\end{equation}
where the $Q_\ijt$ and $Q_\kt$ are the charges of the
emitter and spectator and $\theta_{\ijt/\kt} = 1\,(-1)$
if they are the final (initial) state.
The $\kappa_{\ijt\kt}$ parameters, which define the
charge correlator for photon emitters, can be chosen
to implement a spectator-assigning scheme but are subject to the
constraint that the sum over all possible spectators adds up to the
correct collinear limit.
Various choices are implemented for the NLO subtraction,
see \cite{Schonherr:2017qcj}.
Their phenomenological impact has been explored in
\cite{Flower:2022iew} and found to be minor.
Hence, in the parton shower, the choice
\begin{equation}
  \kappa_{\ijt\kt} = \begin{cases}
    -\frac{1}{N_\text{spec}} & \text{for } Q_\kt \neq 0 \\
    0 & \text{for } Q_\kt = 0,
  \end{cases}
\end{equation}
where $N_\text{spec}$ is the number of charged particles in the process,
has been made.

We use the dipole indices illustrated in Fig.\ \ref{fig:methods:sf:dips}
in the definitions of the splitting functions and other variables.
These differ for each dipole type, and we use the final-final 
notation for a general dipole.

The general form of the dipole splitting functions is
\begin{equation} \label{eq:app:dipoleGen}
  \mr{D}^\mr{A}_{ij,k} = -\frac{1}{2p_i p_j}
  \,\mf{Q}_{\ijt\kt}^2 \,\langle \dots, \ijt, \dots, \kt, \dots
  | \mf{V}_{ij,k} | \dots, \ijt, \dots, \kt, \dots \rangle,
\end{equation}
where $|\dots,\ijt,\dots,\kt,\dots\rangle$ is a vector in 
colour and helicity space, and thus the product indicates a 
squared matrix element, summed over final-state colours and spins.

The parton shower splitting functions are then the spin-averaged 
forms of $\mr{D}^\mr{A}_{ij,k}$, in 4 dimensions, divided by
the Born contribution.
\begin{equation}
  \mr{S}_{\ijt (\kt)\to i j (k)} = -\frac{1}{2p_i p_j} \,\mf{Q}_{\ijt\kt}^2
  \frac{\langle s | \mathbf{V}_{ij,k} | s \rangle|_{\epsilon=0}}
  {\mr{B}},
\end{equation}
in terms of the splitting variable $y$ and light-cone momentum
fraction $z$. For each dipole type, these variables are
defined in table \ref{table:methods:yz}.
The remaining splitting functions can be obtained by
interchanging $i\leftrightarrow j$, and simultaneously
$z \leftrightarrow 1-z$.

\paragraph*{Final-final.} For a final-state splitter $\ijt$ and
a final-state spectator $\kt$,
\begin{equation} \label{eq:app:FF:SFs}
  \begin{split}
    \mr{S}_{f_\ijt(\kt)\to f_i\gamma_j(k)}
    \,=\;
    \mr{S}_{\bar{f}_\ijt(\kt)\to \bar{f}_i\gamma_j(k)}
    \,=&\;
      -\frac{8 \pi \,\alpha\,Q_\ijt^2}{(p_i + p_j)^2-m_{\ijt}^2}\,\mf{Q}_{\ijt\kt}^2\,
      \left[\frac{2}{1-z+zy}
      -\frac{\tilde{v}}{v}\left(1+z+\frac{m_i^2}{p_ip_j}\right) \right] \\
    \mr{S}_{\gamma_\ijt(\kt)\to f_i\bar{f}_j(k)}
    \,=&\;
      -\frac{8 \pi \,\alpha\,Q_i^2}{(p_i + p_j)^2}\,\mf{Q}_{\ijt\kt}^2\,
      \left[
        1-2(z(1-z)-z_+ z_-) \vphantom{\frac{m_i^2}{p_ip_j}}
      \right].
  \end{split}
\end{equation}
The relative velocities $v$ and $\tilde{v}$ are introduced to facilitate the
analytic integration for the case of massive partons, and the ratio reduces 
to unity for massless particles. 
$z_\pm$ are the phase space boundaries. Explicit expressions for these are 
given in the \Sherpa parton shower literature, e.g. ref.~\cite{Schumann:2007mg}.


\paragraph*{Final-initial.} For a
final-state splitter $\ijt$ and a massless
initial-state spectator $\at$,
\begin{equation} \label{eq:app:FI:SFs}
  \begin{split}
    \mr{S}_{f_\ijt(\at)\to f_i\gamma_j(a)}
    \,=\;
    \mr{S}_{\bar{f}_\ijt(\at)\to \bar{f}_i\gamma_j(a)}
    \,=&\;
      -\frac{8\pi\,\alpha\,Q_\ijt^2}{(p_i + p_j)^2-m_{\ijt}^2}\,\mf{Q}_{\ijt\at}^2\,
      \left[\frac{2z}{1-z+y} + (1-z) - \frac{m_i^2}{p_i p_j}\right] \\
    \mr{S}_{\gamma_\ijt(\at)\to f_i\bar{f}_j(a)}
    \,=&\;
      -\frac{8\pi\,\alpha\,Q_i^2}{(p_i + p_j)^2}\,\mf{Q}_{\ijt\at}^2\,
      \left[
        1-2(z_+ - z)(z-z_-)
      \right],
  \end{split}
\end{equation}
where as before, $y$ is the splitting variable defined in Tab.\
\ref{table:methods:yz}.

\paragraph*{Initial-final.} For a massless
initial-state splitter $\ajt$ and a final-state spectator $\kt$,
the IF splitting functions are 
\begin{equation} \label{eq:app:IF:SFs}
  \begin{split}
    \mr{S}_{f_\ajt(\kt)\to f_a\gamma_j(k)}
    \,=\;
    \mr{S}_{\bar{f}_\ajt(\kt)\to \bar{f}_a\gamma_j(k)}
    \,=&\;
      -\frac{8 \pi \,\alpha\,Q_\ajt^2}{2 p_a p_j} \,\frac{1}{z}\,\mf{Q}_{\ajt\kt}^2\,
      \left[\frac{2z}{1-z+y} +(1-z)\right] \\
    \mr{S}_{\gamma_\ajt(\kt)\to f_a\bar{f}_j(k)}
    \,=&\;
      -\frac{8 \pi \,\alpha\,Q_a^2}{2 p_a p_j} \,\frac{1}{z}\,\mf{Q}_{\ajt\kt}^2\,
      \left[
        1-2z(1-z)
      \right].
  \end{split}
\end{equation}
The splitting variables $z$ and $y$ are given in table \ref{table:methods:yz},
taking care to note that the dipole labels are different from
the FI case, see Fig.\ \ref{fig:methods:sf:dips}.


\paragraph*{Initial-initial.} For a massless initial-state splitter
$\ajt$ and a massless initial-state spectator $\bt$,
\begin{equation} \label{eq:app:II:SFs}
  \begin{split}
    \mr{S}_{f_\ajt(\bt)\to f_a\gamma_j(b)}
    \,=\;
    \mr{S}_{\bar{f}_\ajt(\bt)\to \bar{f}_a\gamma_j(b)}
    \,=&\;
    -\frac{8\pi\,\alpha\,Q_\ajt^2}{2p_a p_j} \,\frac{1}{z} \,\mf{Q}_{\ajt\bt}^2\,
      \left[\frac{2(z+y)}{1-z} +(1-z-y)\right] \\
    \mr{S}_{\gamma_\ajt(\bt)\to f_a\bar{f}_j(b)}
    \,=&\;
      -\frac{8\pi\,\alpha\,Q_a^2}{2p_a p_j} \,\frac{1}{z} \,\mf{Q}_{\ajt\bt}^2\,
      \left[
        1-2(z+y)(1-z-y)
      \right].
  \end{split}
\end{equation}

These splitting functions can be extended to take into account massive
initial-state particles, provided that the corresponding subtraction
is implemented for \MCatNLO and that the PDF or structure function,
as well as the corresponding quasi-collinear factorisation used,
is suitable for use with massive partons \cite{Krauss:2017wmx,Forte:2019hjc,Caola:2020xup,Gauld:2021zmq,Harz:2022ipe,
Guzzi:2024can}.

\section{Usage and settings} \label{app:settings}

In this appendix we list the available settings which steer the calculation of the effects described in this paper.
They will be available from a future version of \Sherpa 3. 
All are correct at the time of publication, however, these are subject to change between versions of \Sherpa, so the reader is encouraged to consult the appropriate version of the manual.

For the QED parton shower, like the QCD parton shower, the scale setter must be set to \texttt{MEPS}. 
Under the scoped setting \texttt{SHOWER}, the following settings can be changed:
\begin{description}
    \item[\texttt{EW\_MODE}] \
      This setting turns on splitting functions for the QED parton shower.
      \begin{itemize}
        \item[\texttt{0}] no QED splittings (default),
        \item[\texttt{1}] all QED splittings,
        \item[\texttt{3}] photon emissions off fermions only.
      \end{itemize}
    \item[\texttt{QED\_FS\_PT2MIN}] \
      The infrared cutoff in $t$, the parton shower ordering variable, 
      for the QED final-state evolution.
      Default is $10^{-8}\,\GeV^2$.
    \item[\texttt{QED\_IS\_PT2MIN}] \
      The infrared cutoff in $t$, the parton shower ordering variable, 
      for the QED initial-state evolution.
      Default is $10^{-6}\,\GeV^2$.
    \item[\texttt{QED\_SPECTATOR\_SCHEME}] \
      This setting controls which charged particles may act as spectators 
      for the emission of a photon off a fermion.
      \begin{itemize}
        \item[\texttt{One}] only the nearest opposite-sign same flavour particle considered (default),
        \item[\texttt{All}] all charged external legs in the event considered.
      \end{itemize}
      The nearest particle is selected by taking the minimum dipole invariant mass.
      All charged particles are always considered as spectators for 
      photon splittings and are given equal weight.
      There is an analogous setting under the \texttt{MCATNLO} scope (default : \texttt{All}).
    \item[\texttt{QED\_ALLOW\_FI}] \
      This setting controls whether to allow initial-final interference 
      in the QED shower. Note that if switched on, 
      \texttt{QED\_SPECTATOR\_SCHEME} must be set to \texttt{All}.
      \begin{itemize}
        \item[\texttt{false}] no initial-final interference (default),
        \item[\texttt{true}] initial-final and final-initial dipoles included.
      \end{itemize}
    \item[\texttt{QED\_ALLOW\_II}] \
      This setting controls whether to allow initial-state particles to 
      participate in the QED parton shower.
      \begin{itemize}
        \item[\texttt{false}] no initial-state QED radiation,
        \item[\texttt{true}] initial-initial dipoles included (default).
      \end{itemize}
      This setting is included to facilitate comparisons with other tools
      and for efficiency improvements for quark initial states.
    \item[\texttt{PDF\_MIN\_X}] \
      This setting controls a threshold on the PDF or structure function ratio in the shower for numerical stability. The default is $10^{-2}$, but the QED initial-state shower for leptons requires it to be set to $10^{-1}$ to resolve softer radiation.
\end{description}

To run \Sherpa at NLO in the EW theory, the \texttt{NLO\_Order} parameter in the 
\texttt{Process} section of the runcard must be set appropriately and the loop provider specified. 
In addition, the subtraction must be set as follows:
\begin{description}
  \item[\texttt{NLO\_SUBTRACTION\_MODE}] \
    This setting controls which IR divergences are subtracted.
    \begin{itemize}
      \item[\texttt{QCD}] only QCD subtraction (default),
      \item[\texttt{QED}] only QED subtraction,
      \item[\texttt{QCD+QED}] all SM IR divergences will be subtracted.
    \end{itemize}
\end{description}
The \MCatNLO is turned on in the usual way, using \texttt{NLO\_MODE: MC@NLO} and the associated settings.
Some settings are read from the \texttt{SHOWER} equivalents but some are set separately for
\MCatNLO, please check the latest manual for details.
\bibliographystyle{amsunsrt_modpp}
\bibliography{journal}
\end{document}